\documentclass[twocolumn,prc,aps,showpacs,amsmath,amssymb,nofootinbib]{revtex4}
\usepackage{graphicx}
 \newcommand\la{\langle}
 \newcommand\ra{\rangle}
 \newcommand\beq{\begin{equation}}
 
 \newcommand\eeq{\end{equation}}
 \newcommand\beqn{\begin{eqnarray}}
 \newcommand\eeqn{\end{eqnarray}}
 \newcommand\GeV{{\rm GeV}}
 
\def\im{\mbox{Im}\,}
\def\re{\mbox{Re}\,}

\def\fm{\,\mbox{fm}}
\def\GeV{\,\mbox{GeV}}
\def\TeV{\,\mbox{TeV}}

\def\lsim{\mathrel{\rlap{\lower4pt\hbox{\hskip1pt$\sim$}}
    \raise1pt\hbox{$<$}}}         
\def\gsim{\mathrel{\rlap{\lower4pt\hbox{\hskip1pt$\sim$}}
    \raise1pt\hbox{$>$}}}         

\def\la{\langle}
\def\ra{\rangle}

\begin{document}

\title{ Suppression vs enhancement of heavy quarkonia in pA collisions}

\author{B. Z. Kopeliovich}
\author{Iv\'an Schmidt}
\author{M. Siddikov}

\affiliation{\centerline{$^1$Departamento de F\'{\i}sica,
Universidad T\'ecnica Federico Santa Mar\'{\i}a; and}
Centro Cient\'ifico-Tecnol\'ogico de Valpara\'iso;
Casilla 110-V, Valpara\'iso, Chile}

\begin{abstract}
We describe production of heavy quarkonia in $pA$ collisions within the dipole approach, assuming dominance of the perturbative color-singlet mechanism (CSM) in the $p_T$-integrated cross section. 
Although accounting for a nonzero heavy $Q$-$\bar Q$ separation is a higher twist correction, usually neglected, we found it to be the dominant source of nuclear effects, significantly exceeding the effects of leading twist gluon shadowing and energy loss. Moreover, this contribution turns out to be the most reliably predicted, relying on the precise measurements of the dipole cross section at HERA.
The nuclear suppression of quarkonia has been anticipated to become stronger with energy,
because the dipole cross section steeply rises. However, the measured nuclear effects remain essentially unchanged within the energy range from RHIC to the LHC.
A novel production mechanism is proposed, which enhances the charmonium yield.  Nuclear effects for the production of $J/\psi$, $\psi(2S)$, $\Upsilon(1S)$ and $\Upsilon(2S)$ are calculated, in agreement with data from RHIC and LHC. The dipole description offers a unique explanation for the observed significant nuclear suppression of $\psi(2S)$ to $J/\psi$ ratio,
related to the nontrivial features of the $\psi(2S)$ wave function.

\end{abstract}



\maketitle

\section{Introduction}

Inelastic interactions of a heavy quark pair propagating through a nucleus is a higher twist effect, $\sim1/m_c^2$, which is therefore usually neglected, while leading twist gluon shadowing is  believed to be the main source of nuclear suppression at high energies.
However,
a considerable nuclear suppression of $J/\psi$ production in $pA$ collisions has been observed in the pioneering measurements 
\cite{serpukhov,na3}, even though the energy range of these experiments was too low to explain the observed nuclear effects by gluon shadowing. These data provided the first evidence for the importance of higher twist effects, which certainly remain essential at higher energies and should contribute to the strong  nuclear suppression observed in $pA$ collisions at Fermilab \cite{e866}, RHIC \cite{phenix-psi} and LHC \cite{alice1,alice2}.
Although higher-twist effects as a possible explanation of the observed nuclear suppression of $J/\psi$ was proposed in \cite{pire}, no numerical evaluation was done.

In what follows for concreteness we consider charmonium production, and mainly $J/\psi$,
unless otherwise stated. However, the developed techniques will be also applied to the calculation of nuclear effects in the production of radial excitations and bottomium states.

Charmonium suppression related to the non-zero size, $r\sim1/m_c$, of a perturbatively produced $\bar cc$ dipole, is a higher twist effect, which vanishes in the limit of high quark masses.
Quantitatively, however, it turns out to be the main contributor to the nuclear effects
in charmonium production observed so far \cite{nontrivial,rhic-lhc}. 
At this point we should emphasise
that this higher twist effect is the best known part of nuclear effects. The dipole cross section has been thoroughly measured in deep-inelastic scattering (DIS) at HERA, as function of the dipole energy and size. Therefore, the higher-twist part of dipole attenuation in nuclear matter,
which is responsible for charm nuclear shadowing,
is pretty well known, and leaves not much room for other mechanisms, when is compared with data \cite{nontrivial,rhic-lhc}. On the other hand, leading twist gluon shadowing, which makes nuclear medium more transparent for dipoles, has been poorly fixed by data so far,  ranging from a very weak \cite{florian1,florian2} up to dramatically strong effect \cite{eps08},
even breaking the unitarity bound \cite{klps}.   

These effects lead to reduction of the $J/\psi$ production rate, while the magnitude of shadowing (both, leading and higher twist), as well as the break-up dipole cross section, steadily rise with energy. Therefore, it looks natural to anticipate a stronger suppression of $J/\psi$ produced in $pA$ collisions at the LHC compared with RHIC, as was predicted in \cite{rhic-lhc,andronic}. However, the measurements \cite{alice1,alice2}, unexpectedly revealed energy independent magnitude of $J/\psi$ suppression, which remains unchanged through the huge energy range between RHIC and LHC. This contradiction creates a serious challenge, because as was mentioned above, the dipole phenomenology is well fixed by HERA data, leaving little freedom in its predictions. In spite of the large uncertainly in the gluon shadowing case, it cannot reduce the problem, because its magnitude also rises with energy. 

Here we identify a novel mechanism, which enhances charmonium production and explains the observed anomalous energy dependence. This mechanism was proposed and employed in \cite{hkz} for the explanation of the EMC experiment puzzling data  \cite{emc} on nuclear photoproduction of $J/\psi$. The observed nuclear enhancement was related to non-Glauber double color-exchange interactions with different bound nucleons.
In fact, multiple color-exchange interactions of a dipole propagating through a nuclear medium
lead to a non-vanishing survival probability of the dipole, and even to an enhancement in
specific channels. This is demonstrated in Sect.~\ref{dipole}, based on the results presented in \ref{alesha}.

Nevertheless, the opacity expansion shows that the mean number of color-exchange interactions is rather small even in heavy nuclei, although it rises with energy. Still, the main contribution to $J/\psi$ production is expected to be provided by the single color-exchange interaction considered in Sect.~\ref{single-step}. The cross section on a nucleon tends to  cancel in the nucleus-to-proton ratio, but
the nuclear attenuation factor depends on the features of the parton ensemble propagating through the nucleus. Therefore the description of $J/\psi$ production in $pp$ collisions is essential within the dipole approach, since it allows to calculate the distribution function of the produced
partons in impact parameter space. The details of the calculations are presented in \ref{pp-app}.
The next term of opacity expansion, the double color-exchange interaction, is described in Sect.~\ref{double-step}. The specific challenge here is the calculation of nuclear attenuation factors for $\bar cc$ pairs in certain color states. We found the correction $R^{(2N)}$ to the nuclear ratio $R_{pA}$ to be significant.

Other nuclear effects are also included in the calculations. Gluon shadowing corrections,  evaluated in Sect.~\ref{glue-shad}, are found to be negligibly small at the RHIC energy, but rather significant at the energy of LHC, especially at forward rapidities. Energy loss corrections are considered and introduced in the calculation in Sect.~\ref{eloss}.
The nonperturbative source of energy loss, related to the energy sharing problem at forward rapidities, occurs on a soft scale and brings major corrections to the nuclear effects, as is described in Sect.~\ref{nonpert}. The perturbative mechanism of energy loss, described in Sect.~\ref{nonpert}, is related to the phenomenon of saturation, which generates a new scale,
the saturation momentum, or nuclear broadening. We found the related energy loss to be a quite weak effect, being strongly suppressed by the ratio of the saturation scale to the quarkonium mass squared. This suppression has been missed in previous evaluations, which grossly overestimated  this effect of energy loss.

Special interest has always been paid to nuclear effects in the production of radial excitations,
considered in Sect.~\ref{2s}.
The quarkonium wave function participating in the convolution with the produced $\bar cc$ wave packet, has a node, which leads to a partial compensation between small and large dipole separations. Nuclear color filtering modifies the convolution and can lead to illuminating effects, as was found in the photoproduction of $\psi(2S)$.  The dynamics of hadroproduction is more involved and we arrived at a stronger suppression of $\psi(2S)$ compared to $J/\psi$.
Nuclear effects in $\psi(2S)$ production is a sensitive test of the dipole description of the production mechanism. It provides a unique explanation of the strong suppression of the $\psi(2S)$ to $J/\psi$ ratio in $pA$ collisions. 

The developed dipole description of nuclear effects in charmonium production can be easily extended to heavier quarkonia. In sect.~\ref{upsilon} we perform calculations for the production of $\Upsilon(1S)$ and $\Upsilon(2S)$, in good accord with available data.

\section{Propagation of \boldmath$\bar cc$ dipoles in nuclear medium}
\label{dipole}

\subsection{Characteristic length scales}\label{time-scales}

Two general amplitudes of $\bar cc$  production at different points separated by longitudinal distance $\Delta z$, have a relative phase shift $\Delta\phi=q_L\Delta z$ in the nuclear rest frame, where
the longitudinal momentum transfer is $q_L=M_{\bar cc}^2/2E_{\bar cc}$. 
Correspondingly, the longitudinal length scale $l_c=1/q_L$, usually called coherence length \cite{brod-mueller,kz91}, reads,
\beq 
l_c={1\over q_L}=\frac{2E_{\bar cc}}{M_{\bar cc}^2}.
\label{1005}
\eeq

If the coherence length  exceeds the nuclear dimension, one cannot localize the coordinate  of the $\bar cc$ pair production, in which case the pair propagates through the whole nucleus.
This regime occurs at the energies of RHIC and LHC (except for large negative rapidities).

The $\bar cc$ dipole produced with small transverse separation $r_T\sim1/m_c$,
expands and eventually forms the charmonium wave function on a much longer length scale, called formation length \cite{brod-mueller,kz91},
\beq
l_f\sim \frac{2E_{\bar cc}}{M_{\psi(2S)}^2-M_{J/\psi}^2}\gg l_c
\label{1010}
\eeq
where the masses in the denominator correspond to the first radial excitation $\psi(2S)$ and the $J/\psi$. This can be interpreted in terms of the uncertainty principle as the time interval required to disentangle between the two hadronic masses, while the originally created $\bar cc$ pair has no certain invariant mass and no wave function.

\subsection{Fluctuating dipoles}\label{frozen}

It is clear that at sufficiently high energies the dipole separation does not fluctuate during propagation through the nucleus due to Lorentz time dilation. In this regime the calculations are significantly simplified, so we intend to figure out the kinematic constraints for employing such a "frozen" regime. 

The evolution of a $\bar cc$ dipole propagating through a medium can be described summing up all possible trajectories of the quarks between the initial and final states. 
The amplitude of dipole propagation between longitudinal coordinates $z_1$ and $z_2$, with initial and final transverse separations $\vec r_1$ and $\vec r_2$ respectively, is given by the matrix element of the Green function 
\beqn
A(z_1,z_2)
\!=\!
\int d^2r_1 d^2r_2 \Psi_f^\dagger(\vec r_2)
G(\vec r_2,z_2;\vec r_1,z_1)\Psi_{in}(\vec r_1),
\label{1100}
\eeqn
where $\Psi_{in}(\vec r_1)$ and $\Psi_f(\vec r_2)$ are the initial and final $\bar cc$ distribution amplitudes respectively.

The Green function satisfies the two-dimensional light-cone equation \cite{kz91,kst1,kst2,klss,kpss-psi},
\beqn
&&i\frac{\partial}{\partial z_2}G\left(z_2,\vec r_2;z_1,\vec r_1\right) =
\nonumber\\ &&
\left[\frac{m_c^2-\Delta_{r}}{2E_{\bar cc}\alpha_c\bar\alpha_c}
+V\left(r,z_2\right)\right]
G\left(z_2,\vec r_2;z_1,\vec r_1\right),
\label{322}
\eeqn
with the boundary condition,
$G\left(z_2,\vec r_2;z_1,\vec r_1\right)_{\Delta z=0} =
\delta(\vec r_2-\vec r_1)$. Here
$\alpha_c$ and $\bar\alpha_c=1-\alpha_c$ are the fractional light-cone momenta of $c$ and $\bar c$ respectively. In what follows we fix $\alpha=1/2$ because the charmonium wave function strongly peaks at this value \cite{kth-psi,klss,kpss-psi}. The real part of the light-cone potential $\re V(r)$ describes  the binding effects, while $\im V(r,z)$ is related to the absorption effects, i.e. multiple inelastic  interactions of the dipole  with the medium.

The goal of this section is to figure out the kinematic range of validity of the "frozen" approximation, which corresponds to the high-energy limit, where the formation length Eq.~(\ref{1010}) is much longer than the path length of the dipole in the medium, $l_f\gg\Delta z$.
In this "frozen" dipole regime the Green function approaches the limit $G(\vec r_2,z_2;\vec r_1,z_1) \Rightarrow \delta(\vec r_1-\vec r_2)$, and correspondingly the 
amplitude Eq.~(\ref{1100}) takes the form,
\beq 
A(z_1,z_2) \Rightarrow A_0(z_1,z_2)=
\int d^2r\, \Psi_f^\dagger(\vec r)
\Psi_{in}(\vec r),
\label{1120}
\eeq

In order to quantify the deviation from the "frozen" approximation we evaluate the ratio,
\beq
\epsilon(x_2,\Delta z)=\frac{\left|A(z_1,z_2)\right|^2}
{\left|A_{0}(z_1,z_2)\right|^2},
\label{1140}
\eeq
where $x_{1,2}$ are the fractional light-cone momenta of the colliding gluons,
$gg\to\bar cc$,
\beq
x_{1,2}= \frac{M_T}{\sqrt{s}}\,e^{\pm y},
\label{1160}
\eeq
Here $M_T=\sqrt{M^2_{\bar cc}+p_T^2}$,  $p_T$ and $y$ are the transverse invariant mass, transverse momentum and rapidity (in the $NN$ collision c.m.) of the produced $\bar cc$ pair, respectively. Notice that the dipole energy in the nuclear rest frame is directly related to the value of $x_2$ ,
\beq
E=\frac{M_T^2}{2m_N x_2}.
\label{1162}
\eeq

Anticipating that the validity of the "frozen" approximation means that the result is not sensitive to the details of the binding potential, we evaluate $\epsilon(x_2,\Delta z)$ in a harmonic oscillator potential model\cite{kz91,kst2}, $\re V(r)=(2\omega^2 m_c^2/E) r^2$, where $\omega=(M_{\psi'}-M_{J/\psi})/2\approx 0.3\GeV$. The imaginary part is related to the absorption rate, $\im V(r,z)=C(x_2)r^2\,n_A(z)/2$, where the nuclear density is assumed to be constant, $n_A=0.15\fm^{-3}$, and
the coefficient $C(x_2)$, which controls the dipole cross section at small dipole separations,
was calculated in \cite{broadening} with the parametrization \cite{gbw} of the dipole cross section. In this case Eq.~(\ref{322}) has analytic solution \cite{kz91,kst2}
\beqn
&&
G(\vec r_2,z_2;\vec r_1,z_1)=\frac{N}{2\pi \sinh(\Omega\Delta z)}
\label{1180}\\ &\times&
\exp\left\{-\frac{N}{2}\left[(\vec r_1^{\,2}+\vec r_2^{\,2})\coth(\Omega\Delta z)-\frac{2\vec r_1\cdot\vec r_2}{\sinh(\Omega\Delta z)}\right]\right\}.
\nonumber
\eeqn
Here
\beqn
N^2 &=& \omega^2m_c^2 -
{i\over4}E\,n_A\,C(x_2);
\nonumber\\
\Omega &=& \frac{4iN}{E}.
\label{1200}
\eeqn

\begin{figure}[htb]
\centerline{
  \scalebox{0.7}{\includegraphics{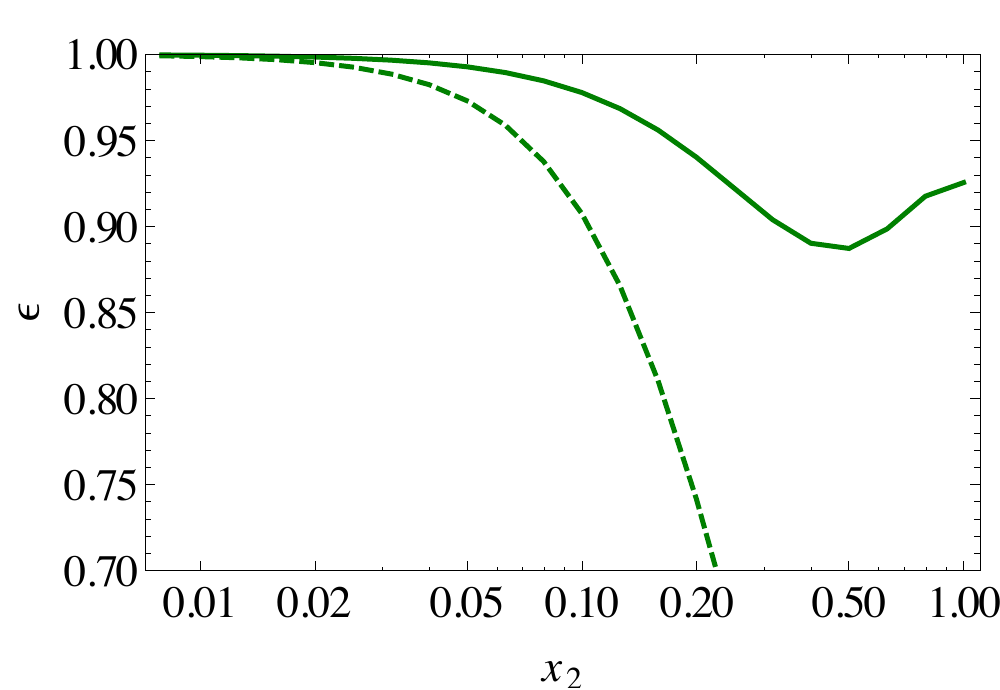}}}
\caption{\label{fig:unfrozen} (Color online) 
Ratio (\ref{1140}) of the dipole propagation probability to the one calculated in the "frozen" approximation. Solid and dashed curves are calculated either with the oscillatory binding potential, or without any potential, respectively. }
 \end{figure}

With this solution we evaluated $\epsilon(x_2,\Delta z)$, Eq.~(\ref{1140}),  fixing $\Delta z=5\fm$ and using the oscillatory $J/\psi$ wave function, as well as the initial distribution function with the mean separation $\la r^2\ra\sim 1/m_c^2$. The results are depicted as function of $x_2$ by a solid curve in Fig.~\ref{fig:unfrozen}.
We see that the "frozen" approximation is valid with a high precision up to rather large values of $x_2\sim 0.1$,
and works reasonably well even at larger $x_2$, matching the Glauber regime. These results confirm
the observation made earlier \cite{kz91}, that the nuclear effects in $J/\psi$ photoproduction remain constant down to quite low energies and are close to the results of the Glauber approximation.

It is instructive to compare this with free $\bar cc$ pair propagation with no binding potential and no absorption. In this case the free Green function is simplified,
\beqn
G(\vec r_2,z_2;\vec r_1,z_1)\Bigr|_{free}&=&\frac{\alpha_c\bar\alpha_c\, E_{\bar cc}}{2i\pi \Delta z}
\label{1220}\\ &\times&
\exp\left[\frac{i\alpha_c\bar\alpha_c\, E_{\bar cc}}{2\Delta z}(\vec r_1-\vec r_2)^2\right].
\nonumber
\eeqn
The corresponding ratio $\epsilon(x_2,\Delta z)$ is depicted by a dotted curve in Fig.~\ref{fig:unfrozen}. We see that even in this extreme case of free expansion the "frozen" approximation is still accurate up to $x_2\sim 0.1$, far more than is needed for the description of available data for $J/\psi$ production at RHIC and LHC. Of course at larger $x_2$ the result significantly deviates from the "frozen" limit, because the quarks freely fly away from each other.

\subsection{Breakup and restoration of colorless dipoles}

According to the conventional wisdom, supported by eikonal-type models, the survival 
probability of a colorless $\bar cc$ dipole propagating through a nuclear medium is exponentially falling with respect to the propagation path length. This is expected to be a result of color-exchange interactions with the surrounding bound nucleons, which break-up the dipole.
However, as is demonstrated below, this is not correct, a high-energy dipole has a finite survival probability even in the limit of full absorption, the so called "black disk" regime \cite{alesha,k-zam}.  

\begin{figure}[htb]
\centerline{
  \scalebox{0.35}{\includegraphics{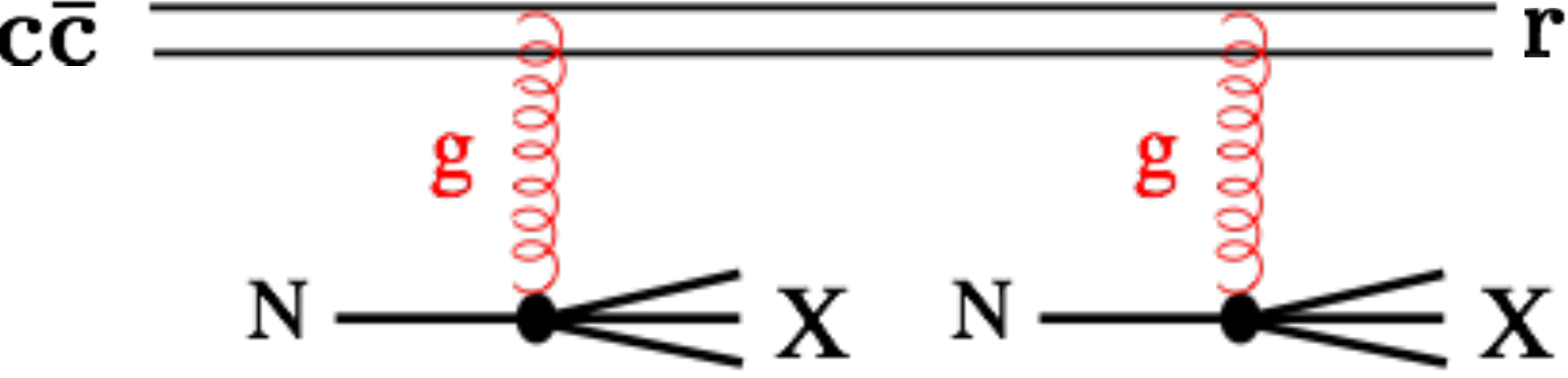}}}
\caption{\label{fig:multiple} (Color online) Multiple color-exchange interaction of a high energy $\bar cc$ pair propagating through a nucleus.}
 \end{figure}

If the dipole energy is sufficiently high, the regime of "frozen"dipoles, described above, remains valid in the medium. Indeed, multiple color-exchange interactions of the dipole with the bound nucleons do not affect the dipole transverse separation, and the interactions only change the color indices  of the quark pair, leading to breakup of the dipole, which becomes colored,
\beq
\bar c^{\,i}c_j+N\to \bar c^{\,k}c_l+X,
\label{120}
\eeq
as is illustrated in Fig.~\ref{fig:multiple}.

 Such interactions also destroy the target nucleons,
so they occur incoherently and should be described in terms of the density matrix $\hbox{}^k_l U^i_j(\vec r;\vec r^{\,\prime};z)$. The
evolution of the density matrix of a high-energy dipole propagating through the nuclear matter is described in \ref{alesha}.
Here we present the results for the probabilities of production of  the final dipole in either color singlet, $S(r)$, or color octet, O(r), states ($\vec r=\vec r^{\,\prime}$).

After propagation through nuclear matter of thickness,
$\Delta T_A=\int_{z_1}^{z} dz'\,n_A(z')$,
where $n_A(z')$ is the nuclear density along the propagation trajectory,
the probability of finding the dipole in a color singlet (S) or octet (O) states,
reads (see derivation in \ref{alesha}),
\beqn
S(r,z)&=&\left[{1\over 9}+{8\over 9}\,e^{-{9\over 8}\sigma_{\bar qq}
(r)\Delta T_A}\right]S_{in}(r);
\nonumber\\
O(r,z)&=&\left[{8\over 9}-{8\over 9}\,e^{-{9\over 8}\sigma_{\bar qq}
(r)\Delta T_A}\right]S_{in}(r).
\label{342}
\eeqn
Here  $S_{in}(r)$ is the size-distribution function of the initial color-singlet dipole;
$\sigma_{\bar qq}(r)$ is the universal dipole-nucleon cross section \cite{zkl}, which depends on  transverse dipole separation and implicitly on the dipole energy or Bjorken $x_2$ (unless specified otherwise).
This cross section is difficult to predict theoretically, but it is well known from phenomenology, fitted to DIS and photoproduction data. A concrete parametrization will be specified later.

Even if the initial state is a color-octet dipole with the size distribution function $O_{in}(r)$, evolution in the medium  may end up with production of either a color singlet, or octet,
\beqn
S(r,z)&=&\left[{1\over 9}-{1\over 9}\,e^{-{9\over 8}\sigma_{\bar qq}
(r)\Delta T_A}\right]O_{in}(r);
\nonumber\\
O(r,z)&=&\left[{8\over 9}+{1\over 9}\,e^{-{9\over 8}\sigma_{\bar qq}
(r)\Delta T_A}\right]O_{in}(r).
\label{382}
\eeqn

We see from (\ref{342}) and (\ref{382}) that for a large number of inelastic collisions of the $\bar cc$ dipole, $\sigma_{\bar cc}(r)\Delta T_A\gg1$,
the probabilities of production of color-singlet and octet states approach the universal values, $1/9$ and $8/9$ respectively, independently of the color structure of the incoming $\bar cc$ pair.
This could be easily anticipated, since both quarks become completely unpolarized in color after multiple rotations in the color space. All possible $9$ color states ($N_c^2$) of the $\bar cc$ are produced with equal probabilities, and only one of them is a singlet, while the 8 others ($N_c^2-1$) are octets.

\subsection{Opacity expansion}

The mean number of inelastic (color-exchange) collisions of a $\bar cc$ dipole of transverse quark separation $r$, propagating through the nucleus, is,
\beq
n^{\bar cc}_{coll}(r,B)=\sigma_{\bar cc}(r,E_{\bar cc})\,T_A(B),
\label{400}
\eeq
where the nuclear thickness function at impact parameter $B$ reads, 
\beq
T_A(B)=\int\limits_{-\infty}^{\infty}dz\,n_A(B,z),
\label{420}
\eeq
and $n_A(B,z)$ is the nuclear density.

For the energy dependence of $\sigma_{\bar cc}(r,E_{\bar cc})$ we rely on parametrizations in the saturated form \cite{gbw,bgbk,amir} for $\sigma_{\bar cc}(r,x)$, fitted to HERA data on the proton structure function $F_2(x,Q^2)$. We are interested in rather low values of $Q^2\sim M_{\bar cc}^2$, for which  even the simple parametrization \cite{gbw} works well \cite{hikt}. 
The value of  target fractional momentum of a target gluon $x_2=e^{-y}\sqrt{(M_{\bar cc}^2+p_T^2)/s}$, controls the magnitude of the dipole cross section.
Here $y$ is the rapidity of the produced $\bar cc$ pair; $p_T$ is its transverse momentum, which is of the order of the mean value, because we are interested in the $p_T$-integrated cross sections.

The dipole cross section steeply rises with energy at small separations,
$\sigma_{\bar qq}(r,x)\sim (1/x)^{0.3}$. At  energy $\sqrt{s}=200\GeV$
and at the measured so far rapidity range $0<y\lsim2$, the mean number of collisions preceding the production of the final colorless $\bar cc$, is $n^{\bar cc}_{coll}\sim
0.05-0.1$. Correspondingly, at energy $\sqrt{s}=5\TeV$
and $0<y\lsim3$, $n^{\bar cc}_{coll}\sim 0.1-0.2$. 

In view of such a small probability of interaction, we keep only the two lowest order terms in the opacity expansion:
(i) single-step direct production \cite{kth-psi,rhic-lhc} of charmonium by the projectile gluon interacting with a bound nucleon, $gN\to \{\bar cc\}_{\psi}X$, with coordinates $(z,\vec B)$, with no preceding or final state interactions;
(ii) a double-step process \cite{hkz}, with the production of a color-octet dipole, $gN\to\{\bar cc\}_8X$, in the first collision, and the final creation of $J/\psi$ in the second collision, $\{\bar cc\}_8+N\to J/\psi+X$.

Correspondingly, the ultimate observable to be calculated, the nucleus-to-proton ratio,
gets contributions from two terms,
\beqn
R^{J/\psi}_{pA}(s,y)&\equiv&\frac{\sigma(pA\to  J/\psi X)}{A\,\sigma(pp\to  J/\psi X)}
\nonumber\\ &=&
R^{(1N)}_{pA}(s,y)+R^{(2N)}_{pA}(s,y)
\label{440}
\eeqn
We assume here that all  cross sections are $p_T$-integrated.

The first term $R^{(1N)}_{pA}$, single-step production, was evaluated for the production of $\chi_2$
in \cite{kth-psi} and for $J/\psi$ in \cite{nontrivial,rhic-lhc}. While this term alone reproduces RHIC data reasonably well, the nuclear suppression predicted  in \cite{rhic-lhc} for the LHC, turned out to be too strong compared with the latest measurements  \cite{alice1,alice2}. Data show that nuclear suppression of $J/\psi$ remains nearly unchanged within the wide energy range from RHIC to LHC. This is impossible for $R^{(1N)}_{pA}$, because the dipole cross section, well constrained by HERA data, rises steeply with energy, leading to a stronger nuclear attenuation of dipoles and smaller values of $R^{(1N)}_{pA}(s,y)$ at higher energies. Therefore, the observed similarity of nuclear effects at RHIC and LHC indicates the onset of a new mechanism, which enhances $J/\psi$ production at LHC energies.
A natural candidate for such a mechanism is the double-step term  in (\ref{440}), which indeed gives a positive contribution, which rises with energy  faster than the single-step term. 

At this point a word of caution is in order. The above estimates for the opacity expansion 
assumed the same interaction  cross section for each of multiple collisions. If, however, the double-step production is dominated by the color-singlet mechanism, the term $R_2$ turns out to be a ratio of different mechanisms. Moreover, gluon radiation in the color-singlet model (CSM)
brings an additional factor $r$ (dipole size) into the amplitude, and then the $r$-dependences of $R_1$ and $R_2$ become similar. Therefore, one can make reliable conclusions about the relative values of the two terms in Eq.~(\ref{440}) only after performing detailed calculations, presented below.

\section{Single-step \boldmath$J/\psi$ production}\label{single-step}

At first glance, if we assume that charmonium is produced on a bound nucleon in the same way as on a free one (see however Sect.~\ref{eloss}), the production cross section on a nucleon should cancel  in the first term $R^{(1N)}$ of the nuclear ratio Eq.~(\ref{440}),
as happens in Glauber-type models. However, attenuation of the projectile and produced partonic ensembles  propagating through the nucleus depends on the mutual transverse separations between the partons, which are controlled by the production mechanism.
Therefore, the nucleon cross section of $J/\psi$ production does not cancel out and affects the nuclear ratio  $R^{(1N)}$, which becomes model dependent.
 
First of all, one should specify the model for $J/\psi$ production, $pp\to J/\psi X$.
Currently the most successful parameter-free description of data has been achieved within the color-singlet model (CSM) proposed in \cite{csm1,csm2}, with further developments and applications in \cite{csm3,csm4,csm5}.
Production of $J/\psi$ is treated in CSM perturbatively, as a result of
glue-glue fusion resulting in production of a colorless $S$-wave $\bar cc$ pair and a gluon
as is illustrated in Fig.~\ref{fig:csm} (left). Gluon radiation allows the $\bar cc$ dipole to have
$S$-wave symmetric wave function (see below).
\begin{figure}[htb]
\centerline{
  \scalebox{0.33}{\includegraphics{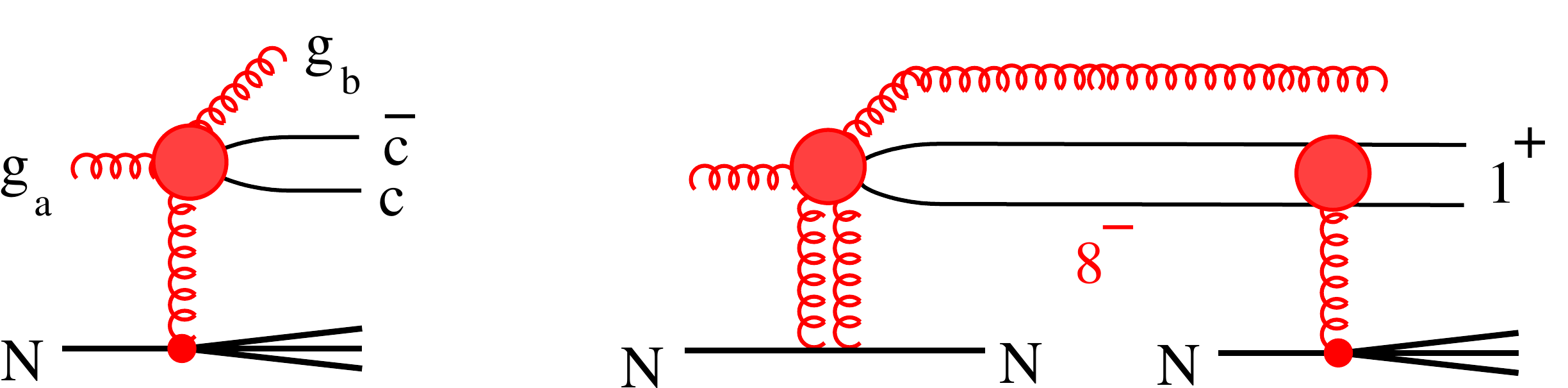}}}
\caption{\label{fig:csm} (Color online)  {\sl Left:} symmetric $1^+$ state production in glue-glue fusion, $gg\to g\{\bar cc\}_{1^+}$. {\sl Right:} diffractive production of color-octet  state $g+N\to g\{\bar cc\}_{8^-}+N$ with subsequent color-exchange transition  $8^-\to 1^{+}$ on another nucleon.}
 \end{figure}

Another popular approach, called color octet model, is based on the non-relativistic QCD effective field theory \cite{nrqcd,raju,Fujii:2013gxa,Ducloue:2015gfa,Ducloue:2016pqr}. The main assumption of the model is that color neutralization occurs via evaporation of soft gluons on a long time scale, of the order of the formation time 
Eq.~(\ref{1010}). Such an unjustified assumption has obvious problems. The initial color-octet $\bar cc$ pair is produced perturbatively at a hard scale $Q^2\sim 4m_c^2$, with no soft gluonic field with frequencies $k_T<m_c$. The laking field is regenerated via perturbative radiation of gluons, making possible $J/\psi$ production in color octet to singlet transition $\{\bar cc\}_{8^-}\to g\{\bar cc\}_{1^+}$, which is a part of the CSM (see details and notations below). 
In this way the $\bar cc$ pair can survive as a color octet and evolve its virtuality down to low scale of the order of the inverse mean $J/\psi$ radius, and then radiate gluons non-perturbatively (color evaporation). However, the probability of scale evolution without gluon radiation, neutralizing the dipole color, is suppressed by a Sudakov-like factor, which is ignored in the color-octet model.

Moreover, the idea of preferable color neutralization at a soft scale, enhanced by a large value of the QCD coupling, does not seem to be correct either. Indeed, according to the Low theorem \cite{low}  the matrix element of a process with soft  radiation is proportional to the process amplitude without radiation, which is impossible for $J/\psi$ production~\footnote{We thank Yuri Dokshitzer for this remark.}. Besides, this model has low predictive power, because it fits the unknown parameters to the data to be explained.
In view of all that, we will consider the color-octet model as a dominant mechanism of $J/\psi$ production. 

Another alternative to the CSM is the possibility of producing $J/\psi$ without gluon radiation, but via $1+2$ gluon fusion, where the two gluons originate either from the beam, or target.  However, evaluation of the cross section \cite{motyka} results in an order of magnitude smaller production rate in comparison to the CSM. We disregard this contribution in what follows.
Nevertheless, a precaution is required for $J/\psi$ production at very forward rapidities, where CSM is suppressed by the shrinking phase space for gluon radiation.

\subsection{Initial state shadowing vs final state attenuation}

As we already discussed in Sect.~\ref{time-scales}, at sufficiently high energies any short time interval is subject to Lorentz time dilation and  becomes long. Even a hard collision, which is characterized by a very short time scale $\tau\sim1/Q$ in its c.m. frame, may last long time (see (\ref{1005})) in the target rest frame, longer than the nucleus dimension.  
In this limit  $J/\psi$ production can be treated as a result of interaction of the $|\bar ccg\ra$ Fock component of the incoming gluon with the whole nucleus. 

Formally one can derive this  adding up the two amplitudes
depicted in Fig.~\ref{fig:csm}.
The first one corresponds to the direct production of the final $S$-wave colorless $\bar cc$ pair symmetric in spacial and spin variable, denoted by $\{1^+\}$.
Another contribution, depicted by the right picture in Fig.~\ref{fig:csm}, contains diffractive  on-mass-shell production of the projectile gluon fluctuation $g\to \bar cc g$, preceding the color-exchange interaction. In order to end up with the production of a $J/\psi$, the $\bar cc$ pair in this fluctuation should be a $P$-wave
color octet  state, asymmetric in spacial-spin variables, which we denote by $\{8^-\}$ \cite{kt-hf}. This color octet pair undergoes color-exchange interactions with the same bound nucleon, as in the first term of the amplitude, and switches to the final colorless $\{1^+\}$ state. 

While the color-exchange interaction occurs on different nucleons {\sl incoherently}, the diffractive production on different nucleons is a {\sl coherent} process.
If the coherence length (inverse longitudinal momentum transfer) is much longer than the nucleus radius, the result is equivalent to interaction of a $|\bar ccg\ra$ fluctuation with the whole nucleus \cite{hkz,ivan-stan}.

Nuclear effect calculations in the CSM have been performed so far in momentum representation \cite{csm1,csm2,csm3,csm4,csm5}, which makes them hardly possible, and in fact this is the reason why the dipole representation for high-energy interactions was first proposed in \cite{zkl}, and extensively used,
in particular for charmonium production off nuclei \cite{hikt2,kth-psi,kt-hf,nontrivial,rhic-lhc,kpss-psi}.
On the other hand, multiple interactions in a nucleus factorize in impact parameter representation, which is then the most appropriate for calculation of the nuclear effects.

\subsubsection{CSM via dipoles: the size distribution} \label{csm}

First of all, one should 
formulate the CSM in terms of dipole interactions. As is explained in detail in Ref.~\cite{kt-hf}, the cross section of the process $g+p\to\bar ccg+X$ is given by the cross section of of the 4-body dipole $|gg\bar cc\ra$, $\sigma_4(\vec r,\vec\rho,\alpha,\alpha_g)$, where $\vec r$ is the $c$-$\bar c$ transverse separation; $\vec\rho$ is the transverse distance between the center of gravity of the $\bar cc$ and the radiated gluon. The second gluon in the 4-body dipole
is the time inverted initial gluon, whose transverse position coincides with the center of gravity of the whole system. The fraction of the light-cone momentum of the initial gluon, carried by the final gluon is $\alpha_g$; and the fractional momenta of $c$ and $\bar c$ inside the produced 
colorless dipole, projected to the $J/\psi$ wave function, are $\alpha$ and $\bar{\alpha}=1-\alpha$ respectively.

Notice that that the mean values of $\la r^2\ra$ and $\la\rho^2\ra$ are controlled
by different mass scales. While the former is related to the heavy quark mass, $r\sim1/m_c$,  the latter is controlled by a semi-hard scale, related to the nonperturbative dynamics. It has been determined by phenomenological analysis of data \cite{kst2,spots}, with fixed $m_g^2\approx 0.5\GeV^2$,  which can be treated as an effective gluon mass squared.
The calculations are significantly simplified, if the small $\la r^2\ra$ is neglected compared with
$\la\rho^2\ra$. Then the color-octet $\bar cc$ pair can be treated as point-like, i.e. is equivalent to a gluon, so the $\sigma_4$ takes the form of a 3-gluon dipole cross section \cite{kt-hf},
\beq
\sigma_4(\rho,\alpha_g)={1\over2}\left[\sigma_{gg}(\rho)+\sigma_{gg}(\alpha_g\rho)+\sigma_{gg}(\bar\alpha_g\rho)\right],
\label{550}
\eeq
where
\beq
\sigma_{gg}(\rho)={9\over4}\,\sigma_{\bar qq}(\rho).
\label{600}
\eeq
We remind that all dipole cross sections depend also implicitly on $x_2$, related to the rapidity $y$ of the produced $\bar cc$,
\beq
x_{1,2}= \frac{\sqrt{M_{ \bar cc}^2+p_T^2}}{\sqrt{s}}\,e^{\pm y_{\bar cc}},
\label{650}
\eeq
where $M_{ \bar cc}$, $p_T$ and $y_{\bar cc}$ are the mass, transverse momentum and rapidity of the $\bar cc$ pair produced in glue-glue fusion, with subsequent decays to $J/\psi$ and gluon. For the sake of simplicity we will associate them with the mass and rapidity of the detected $J/\psi$, unless specified otherwise.


For further  calculations we need to make a choice of parametrization of the dipole cross section $\sigma_{\bar qq}(r,x_2)$. Hereafter we  rely on the parametrization  \cite{bgbk}  fitted to HERA data \footnote{More recent analyses, which also include impact parameter dependence of the elastic dipole amplitude are now available \cite{amir}. For our puposes a $b$-integrated cross section is sufficient},
\beq
\sigma_{\bar qq}(r,x_2) = \sigma_0\left\{
1-\exp\left[\frac{\pi^2 r^2 \alpha_s(\mu^2)x_2g(x_2)}{3\sigma_0}\right]\right\},
\label{665}
\eeq
where the parameter $\sigma_0$ and the scale $\mu^2$ are defined in \cite{bgbk}.

The presence of the gluon density in (\ref{665}) shows that this parametrization corresponds to the Pomeron contribution to the dipole cross section. This is the reason why it describes well the DIS data only at sufficiently small $x_2<0.01$  \cite{bgbk}. At larger $x_2$ the Reggeon contribution, which corresponds to valence quarks in $F_2(x,Q^2)$, increases, and the Pomeron alone fails to describe data.
This problem, however, is relevant only for light quarks, which dominate in the $F_2(x_2,Q^2)$ measured at HERA. 
For $\bar cc$ 
dipoles the Reggeon term, corresponding to valence $\bar qq$ exchanges,  is suppressed by the OZI rule \cite{ozi1,ozi2,ozi3}, which suppresses valence charm component in the proton. Smallness of such a component (intrinsic charm \cite{stan1,stan2}) is confirmed by data \cite{pumplin}, so it
can be neglected.

For $\bar cc$ 
dipoles the Reggeon term is suppressed by the OZI rule and can be neglected. 
\cite{ozi1,ozi2,ozi3}
Thus, the parametrization 
Eq.~(\ref{665}) for $\bar cc$ dipoles can be safely extended up to $x_2\sim 0.1$, where
the coherence length Eq.~(\ref{1005}) shrinks down to the nucleon size.

The cross section of $J/\psi$ production is derived in \ref{pp-app} and is given by Eq.~(\ref{b600}). Since the amplitude contains the projection to the $J/\psi$ wave function,
the cross section contains integrations over $\vec r$ and $\vec r^{\,\prime}$. On the other hand, the radiated gluon is not registered, and integration over its transverse momentum produces a delta function $\delta(\vec\rho-\vec\rho\,')$. Therefore 
the size distribution function $W(\vec\rho,\vec r,\vec r^{\,\prime})$ depends on only three variables. We normalize this function to unity,  and
relate it to the $pp$ differential cross section of charmonium production, presented in Eqs.~(\ref{b500})-(\ref{b600}),
\beq
W(\vec\rho,\vec r,\vec r^{\,\prime})=\frac{d\sigma_{pp}^{J/\psi}}
{dy\, d^2\rho\, d^2r\, d^2r^\prime}
\left[\frac{d\sigma_{pp}^{J/\psi}}{dy}\right]^{-1},
\label{800}
\eeq
This distribution also depends implicitly on $x_2$.

\subsection{Nuclear effects}

Now we are in a position to predict the nuclear effects, \beqn
R^{(1N)}_{pA}(s,y)&=&\int d^2B\int\limits_{-\infty}^{\infty} dz\,n_A(B,z)
\int d^2\rho\,d^2r\,d^2r^\prime
\nonumber\\ &\times&
W(\vec\rho,\vec r,\vec r^{\,\prime})\,
S_A^{(1N)}(B,z_1,\vec\rho,\vec r,\vec r^{\,\prime})
\label{700}
\eeqn
 Here $\vec B$ is the impact parameter of the $pA$ collision; $z$ is the longitudinal coordinate of the incoherent color-exchange interaction, which leads to the production of a colorless $S$-wave $\bar cc$ dipole, projected to the $J/\psi$ wave function. The nuclear suppression factor $S_A^{(1N)}$ includes
 shadowing due to  reduction of the $\bar ccg$ flux at $z^\prime<z$ and attenuation of the produced colorless $\bar cc$ dipole at $z^\prime>z$,
\beqn
S_A^{(1N)}(B,z_1,\vec\rho,\vec r,\vec r^{\,\prime}) &=&
\exp\left[ - \sigma_4(\rho,\alpha_g)\,T_-(B,z)\right]
\label{750} \\ &\times&
\exp\left[-\Sigma_1(\vec r,\vec r^{\,\prime})\,T_+(B,z)\right],
\nonumber
\eeqn
where $\sigma_4(\rho,\alpha_g)$ is given by Eq.~(\ref{550}), and $\Sigma_1(\vec r,\vec r^{\,\prime})=[\sigma_{\bar qq}(r)+\sigma_{\bar qq}(r^\prime)]/2$, by Eq.~(\ref{a320}).
 The nuclear thickness functions, $T_-(B,z)$ and $T_+(B,z)$,
 which correspond to the propagation of the projectile $g\bar cc$ fluctuation up to the point $(\vec B,z)$ and the propagation of the produced $\bar cc$ dipole afterwards, respectively.
 \beqn
 T_-(B,z)&=&\int\limits_{-\infty}^z dz'\,n_A(B,z');
 \nonumber\\
 T_+(B,z)&=&\int\limits_z^{\infty} dz'\,n_A(B,z').
 \label{760}
 \eeqn
Apparently, $T_-(B,z)+T_+(B,z)=
T_A(B)$, the full thickness function given by (\ref{420}).

\begin{figure}[htb]
\centerline{
  \scalebox{0.65}{\includegraphics{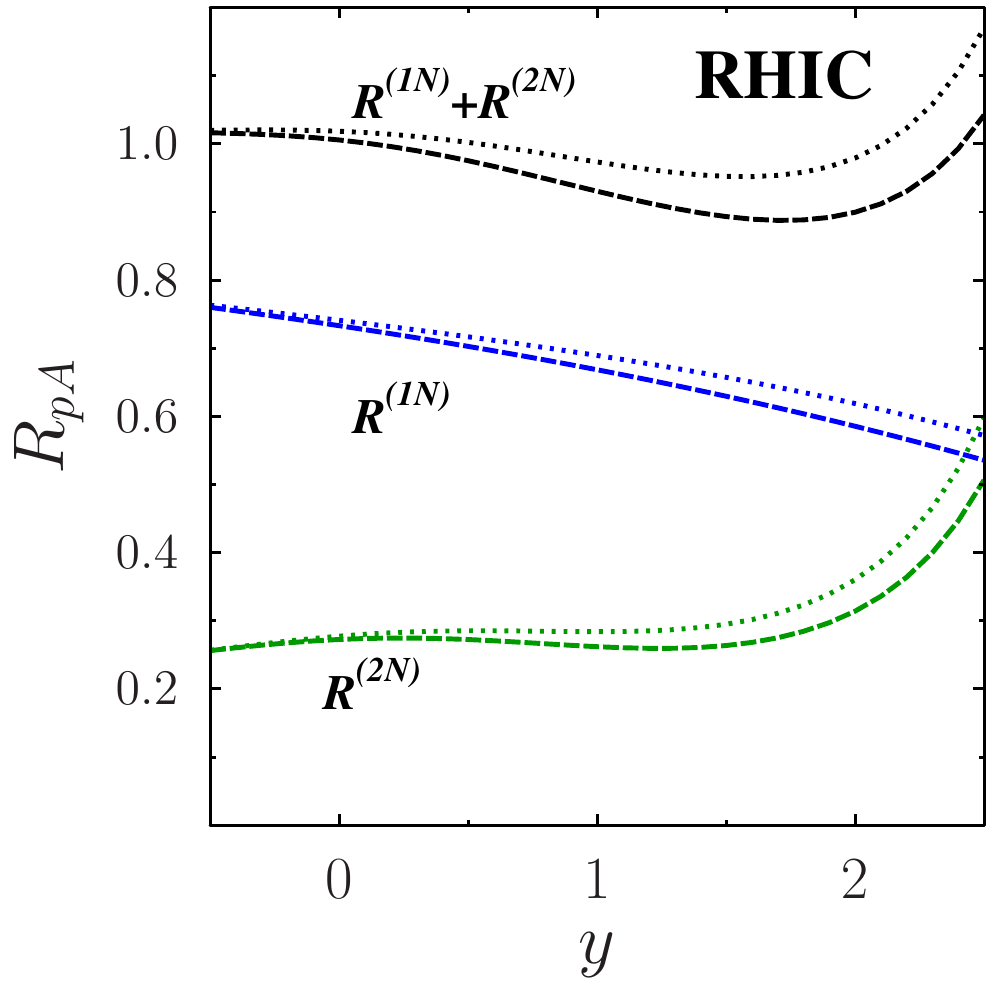}}}
\caption{\label{fig:rhic-glue} (Color online)  From bottom to top, the terms $R^{2N}$, $R^{1N}$ and their sum, Eq.~(\ref{440}), for $p$-$Au$ collisions at $\sqrt{s}=200\GeV$.  Dotted and dashed curves present calculations without and with gluon shadowing corrections respectively.
}
 \end{figure}

\begin{figure}[htb]
\centerline{
  \scalebox{0.65}{\includegraphics{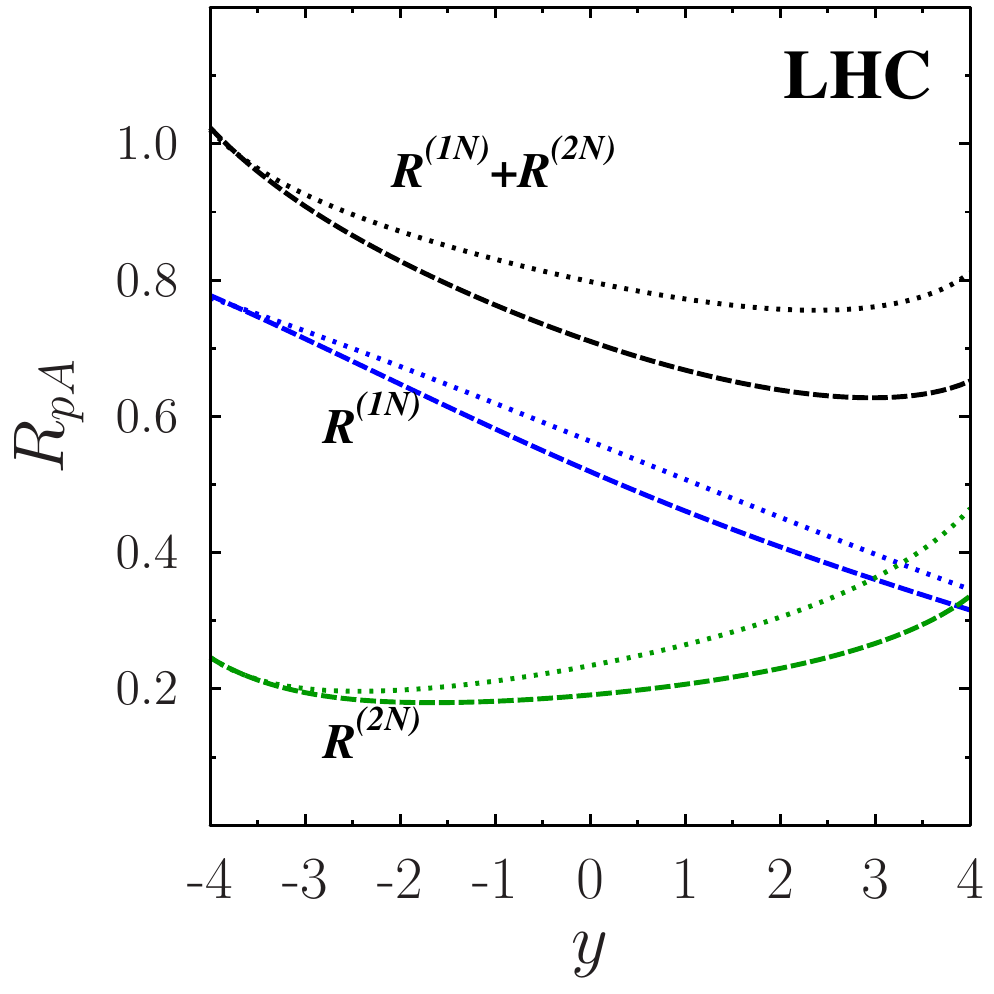}}}
\caption{\label{fig:lhc-glue} (Color online) The same as in Fig.~\ref{fig:rhic-full}, but for $p$-$Pb$ collisions at $\sqrt{s}=5000\GeV$.
}
 \end{figure}

Now we can calculate the single-step term Eq.~(\ref{700}) and the results at $\sqrt{s}=200$ and $5000\GeV$ are plotted in Figs.~\ref{fig:rhic-glue} and \ref{fig:lhc-glue} by blue curves labelled as $R^{1N}$.

These results are close to the first simplified calculations done in \cite{nontrivial,rhic-lhc},
which agreed reasonably well with data \cite{phenix-psi} at $\sqrt{s}=200\GeV$,  but 
grossly under-predicted the ratio $R_{pA}$ at the energy of LHC \cite{alice1,alice2}. This fact was already
highlighted in \cite{andronic}.

Notice that such a contradiction with the observed energy dependence of the nuclear ratio is not a simple failure of a concrete model, but discloses a deeper puzzle. The dipole cross section is well constrained by  precise DIS data from HERA. It is  known to steeply rise with $1/x$, therefore the magnitude of nuclear attenuation of dipoles must rise with energy. This expectation is beyond the details of a particular model, and cannot be easily changed. The observed similarity of nuclear suppression at both RHIC and LHC energies should be treated as an indication of a new mechanism of $J/\psi$ production in nuclei, for which a natural candidate is the second term in Eq.~(\ref{440}).

\section{Double-step production}\label{double-step}

The second term in Eq.~(\ref{440}) is given by,
\beq
R^{(2N)}_{pA}(s,y)=
\frac{\sigma^{(2N)}(pA\to J/\psi X)}
{A\,\sigma(pp\to J/\psi X)},
\label{500}
\eeq
where the double-step contribution to the numerator is illustrated in Fig.~\ref{fig:multiple}.
Summing over final states one arrives at the cross section, expressed in terms of the density matrix, as is described in \ref{alesha}.


The first color-exchange interaction, $g+N\to\bar cc+X$, can result in the production of a $\bar cc$ pair in three different
states (at leading order):
(i) antisymmetric relative to permutations of space and spin variables, 
 color singlet $\{1^-\}$ or color octet $\{8^-\}$states; (ii) symmetric in spacial-spin variables, color octet state $\{8^+\}$. The notations used here are from \cite{kt-hf}

Assuming that the finally produced state after the second interaction is a colorless $S$-wave $\bar cc$ dipole \{$1^+$\}, the intermediate $\bar cc$ pair, between the first and second collisions, must be a $P$-wave $\{8^-\}$ state. The first collision cross section has the form \cite{kth-psi,kt-hf},
\beqn
&&\sigma(gp\to (\bar cc)_{\{8^-\}}X) =
\sum\limits_{\mu,\bar\mu}\,\int\limits_0^1 
d\alpha d\alpha^\prime\int d^2r d^2r^\prime\, 
\nonumber\\ &\times&
{\Phi_g^{\mu\bar\mu}(\vec r,\alpha)}^\dagger
\Phi_g^{\mu\bar\mu}(\vec r^{\,\prime},\alpha^\prime)\,
\Sigma_{g\to\{8^-\}}(\vec r,\vec r^{\,\prime},\alpha,\alpha^\prime),
\label{555}
\eeqn
where $\Phi_g^{\mu\bar\mu}(\vec r,\alpha)$ is the light-cone distribution function of $\bar cc$ incoming gluon, defined in (\ref{350});
\beq
\Sigma_{g\to\{8^-\}}
\approx
{5\over8}\left[\sigma_{\bar qq}\left(\frac{\vec r+\vec r^{\,\prime}}{2}\right) -
\sigma_{\bar qq}\left(\frac{\vec r-\vec r^{\,\prime}}{2}\right)
\right].
\label{605}
\eeq
We fixed here $\alpha=\alpha^\prime=1/2$, because these  
values are strongly enhanced by the projection into the charmonium wave function \cite{kth-psi,klss}.

The second interaction, $\{\bar cc\}_{\{8^-\}}N\to J/\psi X$, is the time reversal
of the usual inelastic (color exchange) interaction, $ J/\psi N\to X$, which is related to the  dipole cross section,
\beqn
\Sigma_{\{8^-\}\to\{1^+\}} 
\approx
{1\over8}\left[\sigma_{\bar qq}\left(\frac{\vec r+\vec r^{\,\prime}}{2}\right) -
\sigma_{\bar qq}\left(\frac{\vec r-\vec r^{\,\prime}}{2}\right)
\right]\!.
\label{655}
\eeqn

Thus, we are in a position to calculate the numerator of the double-scattering term Eq.~(\ref{500}) as,
\begin{widetext}
\beqn
\frac{d\sigma^{(2N)}(pA\to J/\psi X)}{dy}&=&
g_N(x_1)\int d^2B\int\limits_{-\infty}^{\infty} dz_1\,n_A(B,z_1)
\int\limits_{z_1}^{\infty} dz_2\,n_A(B,z_2)
\int\limits_0^1 d\alpha d\alpha^\prime
\int d^2r d^2r^\prime\,
\nonumber\\ &\times&
\Psi^\dagger_{ J/\psi}(\vec r,\alpha)
\left\la 1M\Biggl|{1\over2}\bar\mu{1\over2}\mu\right\ra
\Phi_{\bar cc}^{\bar\mu\mu}(\vec r,\alpha)
\left[
\Psi^\dagger_{ J/\psi}(\vec r^{\,\prime},\alpha^\prime)
\left\la 1m\Biggl|{1\over2}\bar\mu{1\over2}\mu\right\ra
\Phi_{\bar cc}^{\bar\mu\mu}(\vec r^{\,\prime},\alpha^\prime)
\right]^*
\nonumber\\ &\times&
\Sigma_{g\to\{8^-\}}(\vec r,\vec r^{\,\prime},\alpha,\alpha^\prime)\,
\Sigma_{\{8^-\}\to\{1^+\}}(\vec r,\vec r^{\,\prime},\alpha,\alpha^\prime)\,
S_A^{(2N)}(B,z_1,z_2,\vec r,\vec r^{\,\prime},\alpha,\alpha^\prime),
\label{720}
\eeqn
\end{widetext}
where the gluon PDF in the beam proton, $g_N(x_1)$, is taken at the scale $Q^2=4m_c^2$.
The wave function of quarkonium is normalized according to
\beq
\int d\alpha_{Q}d^{2}r_{Q}\,\left|\Psi_{J/\psi}\left(\alpha_{Q},\,\vec{r}_{Q}\right)\right|^{2}=1.
\label{722}
\eeq
For evaluations, we rely on the LC charmonium 
wave function obtained with the Cornell potential~\cite{cornell1,cornell2}
and boosted to another frame following the procedure developed in~\cite{klps}.

The $\bar cc$ light-cone distribution function is convoluted in (\ref{720}) including 
the Clebsch-Gordan coefficient $\left\langle 1M\left|\frac{1}{2}\bar{\mu}\frac{1}{2}\mu\right.\right\rangle$, and $M$ is the spin $z$-projection.
The nuclear suppression factor $S_A^{(2N)}$ is presented below.

\subsection{The nuclear suppression factor}\label{S2}

This factor gets contributions from different
parts of the dipole path through the nucleus (see Fig.~\ref{fig:multiple}) :
(i) prior the first collision at longitudinal coordinate $z_1$ and production of the color-octet, P-wave $\bar cc$ pair $\{8^-\}$;
(ii) attenuation of the produced $\{\bar cc\}_{8^-}$ pair on the path from $z_1$ up to the next color-exchange interaction at $z_2$;
(iii) attenuation of the produced colorless dipole $\{\bar cc\}_{1^+}$on its way out of the nucleus.
Correspondingly, the nuclear suppression can be presented as a product of three factors,
\beq
S_A^{(2N)}=
S_1^{z<z_1}\,S_2^{z_1<z<z_2}\,S_3^{z>z_2}.
\label{770}
\eeq

The first factor $S_1(z<z_1)$ has the meaning of shadowing, namely the competing probabilities of the process $g\to\bar cc$ to occur on different bound nucleons, which
reduce the gluon flux \cite{piller,kt-hf},
\beq
S_1^{z<z_1}=
\exp\left[-\Sigma_3(\vec r,\vec r^{\,\prime},\alpha,\alpha^\prime)
T_-(B,z_1)\right],
\label{805}
\eeq
where $\Sigma_3=\left[\sigma_3(r,\alpha)+\sigma_3(r,^\prime\alpha^\prime)\right]/2$, and
\beq
\sigma_3(r,\alpha) = {9\over8}\left[\sigma_{\bar qq}(\alpha r)+\sigma_{\bar qq}(\bar\alpha r)\right]-{1\over8}\sigma_{\bar qq}(r).
\label{855}
\eeq
The cross section $\sigma_3(r,\alpha)$ controlling the suppression, is the total cross section of a 3-body dipole 
($g\bar cc$), responsible for the inclusive production process $gN\to\bar ccX$
\cite{piller,kt-hf}.

The second factor in (\ref{770}) can be treated as the survival probability
of the produced $(\bar cc)_{\{8^-\}}$ pair propagating through the medium.
Its attenuation is controlled by only a part of the cross section $\Sigma_8(\vec r,\vec r^{\,\prime})$ introduced in (\ref{a320}). While the diagonal transitions  $\{\bar cc\}_{\{8^-\}}\to\{\bar cc\}_{\{8^-\}}$ do not affect the final result,  the other channels, such as transitions of  $\{\bar cc\}_{\{8^-\}}$ to a
singlet  $\{\bar cc\}_{1^+}$, or to a color octet S-wave  $\{\bar cc\}_{\{8^+\}}$, eliminate further possibilities of production of $ J/\psi$ at $z=z_2$. Summing up the cross sections of the last two channels, we arrive at the second suppression factor in (\ref{770}),
\beq
S_2^{z_1<z<z_2}=\exp\biggl[-\Sigma_{\{8^-\}}(\vec r,\vec r^{\,\prime},\alpha,\alpha^\prime)\,
T_{12}(B,z_1,z_2)\biggr],
\label{900}
\eeq
where $T_{12}(B,z_1,z_2)=T_-(B,z_2)-T_-(B,z_1)$, and
\beqn
\Sigma_{\{8^-\}}&=&
{7\over32}\biggl[\sigma_{\bar qq}(\alpha\vec r+\bar\alpha^\prime\vec r^{\,\prime})+
\sigma_{\bar qq}(\bar\alpha\vec r+\alpha^\prime\vec r^{\,\prime})
\nonumber\\ &-&
\sigma_{\bar qq}(\alpha\vec r-\alpha^\prime\vec r^{\,\prime})-
\sigma_{\bar qq}(\bar\alpha\vec r-\bar\alpha^\prime\vec r^{\,\prime})\biggr]
\nonumber\\ &\approx&
{7\over16}\left[\sigma_{\bar qq}\left(\frac{\vec r+\vec r^{\,\prime}}{2}\right)-
\sigma_{\bar qq}\left(\frac{\vec r-\vec r^{\,\prime}}{2}\right)\right].
\label{950}
\eeqn
In the last line we again employ the approximation $\alpha=\alpha^\prime=1/2$, for the sake of simplicity.

The last factor in (\ref{750}) has a rather obvious form,
\beq
S_3^{z>z_2}=
\exp\left[-\Sigma_1(\vec r,\vec r^{\,\prime},\alpha,\alpha^\prime)\,
T_+(B,z_2)\right],
\label{1000}
\eeq
where $\Sigma_1(\vec r,\vec r^{\,\prime})$, is given by Eq.~(\ref{a320}).

Notice that the $z$-dependent part of (\ref{700}) can be integrated analytically,
\beqn
&&\int\limits_{-\infty}^{\infty} dz_1\,n_A(B,z_1)
\int\limits_{z_1}^{\infty} dz_2\,n_A(B,z_2)
\nonumber\\ &\times&
S^{(2N)}(B,z_1,z_2,\vec r,\vec r^{\,\prime},\alpha,\alpha^\prime)
\nonumber\\ &=&
\frac{1-e^{-\Omega_{2}T_A(B)}}{\Omega_{1}\Omega_{2}}-†
\frac{1-e^{-\Omega_{3}T_A(B)}}{\Omega_{1}\Omega_{3}},
\label{1050}
\eeqn
where we introduced the short-hand notations,
$\Omega_{1}=\Sigma_3-\Sigma_{\{8^-\}}$;
$\Omega_{2}=\Sigma_{\{8^-\}}-\Sigma_1$;
$\Omega_{3}=\Sigma_3-\Sigma_{1}$.

\subsection{The \boldmath$pp$ reference}\label{pp}

In our calculation of $R^{(1N)}_{pA}$ for the single-step mechanism, we 
assumed that the same CSM model dominates both the numerator and denominator,
and therefore they have nearly identical functional forms, except for the nuclear suppression factor 
and some corrections discussed below.
So the reference $pp$ cross section nearly cancels.

The  double-step term $R^{(2N)}_{pA}$ evaluation is more peculiar, because the numerator and denominator originate from different mechanisms and have distinct functional forms. While the former, given by Eq.~(\ref{700}), is calculated directly based on the well developed dipole phenomenology,
the latter depends on the choice of a model for inclusive $J/\psi$ production (see Sect.~\ref{single-step}) and is assumed here to be dominated by CSM. Thus, the denominator of $R^{(2N)}_{pA}$ Eq.~(\ref{500}) has a rather wide theoretical uncertainty band, which is related to the accuracy of the CSM, and possibility of other missed contributions (such as three-gluon fusion \cite{motyka}, certainly important at very forward/backward rapidities).

The least model dependent way to treat the denominator of  (\ref{500}) would be to take it directly from a fit to experimental data for $pp\to J/\psi X$, available  within certain kinematic domains. We rely on our evaluations of the $pp\to J/\psi X$ cross section, performed within
the dipole version of the CSM in \ref{pp-app}. The results, compared with data in Figs.~\ref{fig:pp-phenix} and \ref{fig:pp-alice}, well reproduce the shape of the $y$-dependence of the cross section, however, slightly underestimate the normalization. At $\sqrt{s}=200\GeV$ we employed the data from \cite{phenix-psi,phenix-pp}.
Lacking experimental results at  $\sqrt{s}=5\TeV$, we interpolated between  data at $\sqrt{s}=2.76\TeV$ and $7\TeV$ \cite{interpolation}. The details are presented in \ref{pp-app} and the results are depicted in Fig.~\ref{fig:pp-alice}.
Since, as we said, data is the most reliable source of information about the $pp$ cross section, we adjusted the normalization of the theoretical curves to fit the data, keeping the shape of the $y$-dependence unchanged.

Now we are in a position to calculate $R^{2N}$, the ratio of the cross section Eq.~(\ref{720}) to the chosen $pp$ reference, and the results at $\sqrt{s}=200$ and $5000\GeV$ are plotted in Figs.~\ref{fig:rhic-full} and \ref{fig:lhc-full} by dashed curves labelled as $R^{2N}$.

\section{Gluon shadowing}
\label{glue-shad}

Leading twist 
gluon shadowing originates in the nuclear rest frame from coherent multiple interactions of the radiated gluons.
It can also be treated as 
the contribution of higher Fock components in the projectile hadron, containing extra gluons, which have a coherence (radiation) time sufficiently long to experience multiple interactions in the nucleus \cite{kst2}. These gluons are complementary to the gluon radiated within the CSM mechanism (Fig.~\ref{fig:csm}).
Unlike quark shadowing, which is known to onset at $x_2\lesssim 0.1$ \cite{krt-shad1},
gluon shadowing needs an order of magnitude smaller $x_2$ to show up \cite{krt-shad2}.
This is controlled by the coherence length of gluon radiation,
\beq
l_c^{g\bar cc} = \frac{P_g}{x_2m_N},
\label{222}
\eeq
which must be longer than the mean free path in nuclear matter.
The factor $P_g\approx 0.1$, evaluated in \cite{krt-shad2}, makes the coherence time of gluon radiation significantly shorter than the Ioffe time for quarks.
This happens due to the enhanced transverse momenta of gluons in hadrons \cite{kst2,spots},
which make the fluctuations containing gluons much heavier.
For the same reason, the mean quark-gluon separation is short, and the magnitude of the leading-twist gluon shadowing turns out to be rather small, even compared with the higher-twist quark shadowing. The weaknees of gluon shadowing, predicted in \cite{kst2}, was confirmed by the NLO analysis of DIS data \cite{florian1,florian2}.

The gluon shadowing suppression factor $R_g(x,Q^2)$, calculated in \cite{kst2}, was  applied to Drell-Yan process in \cite{jkrt-dy}, and to heavy flavor production in \cite{kt-hf}, where one can find the details of the calculations. 
This factor suppresses $J/\psi$ production on nuclei as well. In our case we include gluon shadowing by reducing the dipole cross section with the shadowing factor $R_g$, which also depends on the nuclear impact parameter $b$. Such a way of incorporation of gluon shadowing can be justified only 
at first order, which corresponds to radiation of a single gluon.  In fact, radiation of two
gluons lead to a quadratically short coherence time compared with Eq.~(\ref{222}) \cite{k-poetic}, too short to cause shadowing at currently available energies.

The terms $R^{1N}$ and $R^{2N}$ in (\ref{440}), with added gluon shadowing corrections 
at $\sqrt{s}=200\GeV$ and $5000\GeV$, are depicted in Figs.~\ref{fig:rhic-glue} and \ref{fig:lhc-glue} respectively. 
The corrections are found to be rather small at the energy of RHIC (due to shortness of the coherence length), but significant at LHC.
Nevertheless, even at the LHC energy gluon shadowing vanishes in the backward hemisphere, towards the minimal rapidity $y\sim-4$ in the kinematical range measured so far, because the coherence length Eq.~(\ref{222}) becomes shorter than the mean spacing between bound nucleons.

\section{Energy loss}\label{eloss}

 \subsection{Nonperturbative energy loss}
 \label{nonpert}

Apparently, multiple soft interactions in the nuclear medium should lead to  
dissipation of energy by the projectile partons, reducing the production rate of $J/\psi$ at large 
Feynman $x_F$, where the restricted phase space of produced $J/\psi$ becomes an issue.  Energy loss was first proposed in \cite{kn-psi} as a mechanism of suppression of the $pA$-to-$pp$ ratio of $J/\psi$ production at large $x_F$, observed in  \cite{serpukhov,na3}.
The rate of energy loss, treated within the string model, was independent of the incoming proton energy \cite{kn-psi}.  Perturbative calculations, performed in the approximation of soft  gluon radiation, confirmed the string model result of energy independent parton energy loss \cite{feri,bh}. 
This, however, could not explain the observed $x_F$ scaling, i.e. similarity of the $x_F$-dependences of nuclear effects in $J/\psi$ production at different energies \cite{serpukhov,na3,e866}.

Nonetheless, later, in \cite{e-loss,boosting}, it was found that the rate of energy loss, either in nonperturbative \cite{e-loss}, or perturbative \cite{boosting} regimes, rises proportional to the incoming energy. This is easily interpreted in terms of Fock-state representation for the light-cone wave function of the incoming hadron.
The probability of giving a significant fraction of the hadron momentum to one parton (soft or hard) is more suppressed in the higher Fock states. Indeed, if one of the participating partons  gets a large momentum fraction $x_1\to1$, all other participants are pushed into 
a small phase space with $x<1-x_1$.
 The measured parton distribution function (PDF) is averaged over different Fock components, and the interaction of these Fock states with the nuclear target changes
their weights, increasing the contribution of higher Fock components, so that the projectile parton distribution becomes softer, i.e. more suppressed at large $x_1\to1$. 
Thus, the projectile proton PDF becomes target-dependent, violating QCD factorization
 at large $x_1$, where the energy sharing (energy loss) problem becomes important \cite{e-loss,boosting}.   Such a beam-target correlation breaks factorization, because it occurs at a low scale. This explains why every process measured so far was found to be nuclear suppressed at large $x_1$ \cite{e-loss}. 

 Glauber multiple hadron-nucleus soft inelastic interactions are not sequential (as is frequently naively believed), but correspond to multi-sheet configurations in the topological $1/N_c$ expansion of QCD for the
inelastic amplitude, i.e.  they are related to simultaneous propagation and interaction in the medium of different projectile partons from a high Fock component of the incoming hadron \cite{gribov1,gribov2,agk,kaidalov,capella}. This leads to the problem of energy sharing between participating partons, which becomes especially severe at large fractional momentum $x_1$ carried by one of them. The associated nuclear suppression was calculated in \cite{e-loss} using the Fock state expansion, weighted by the interaction with the target, corresponding to the Glauber model. The suppression factor $S(x_1)$ for each additional topological sheet was evaluated in \cite{kaidalov, capella} relying on Regge phenomenology, and in \cite{e-loss} treating it as a rapidity gap survival probability. Both approaches led to the same result: at $x_1\to1$ suppression increases as $S\propto(1-x_1)$.
We apply here the model for energy loss developed in \cite{e-loss} , in order to correct the nuclear ratio (\ref{440}).

 \subsection{Perturbative energy loss}
 \label{pert}
 
 Another source of nuclear modification of the projectile gluon distribution is an increased hard scale. Indeed, if in $pp$ collision the gluon distribution is taken at the scale $Q^2=4m_c^2$, a nuclear target generates another scale, known as saturation scale $Q_s^2$.
So the effective scale of the process increases, $Q_{eff}^2=4m_c^2+Q_s^2$. This 
follows naturally from the interpretation of saturation in the rest frame of the nucleus,
which is related to broadening of the  transverse momentum of a gluon propagating through the nucleus \cite{broadening}, 
\beq
Q_s^2(B,x_2)=\Delta p_T^2=T_A(B)\,{9\over4}\,
\vec\nabla^2\sigma_{\bar qq}(r,x_2)\biggr|_{r=0}.
\label{1300}
\eeq 
We employ the dipole description of broadening \cite{jkt}, and  for the saturation scale rely on the results of \cite{broadening}. This result is based on the approximation of Bethe-Heitler regime of gluon radiation in multiple interactions, neglecting interferences of gluons radiated in collisions with different nucleons. Effects of coherence cause deviations from Eq.~(\ref{1300}),
however according to the discussion in Sect. V effects of coherence in gluon radiation are small even at the energies of LHC. Therefore, in what follows we employ the approximate effective scale $Q_{eff}^2=4m_c^2+Q_s^2$ for numerical evaluations.

Notice that broadening of the  transverse momentum of a gluon propagating through the nucleus is equivalent to the effect of saturation in the $k_T$-dependent PDF of the nucleus in its infinite-momentum frame \cite{saturation}.

\begin{figure}[htb]
\centerline{
  \scalebox{0.6}{\includegraphics{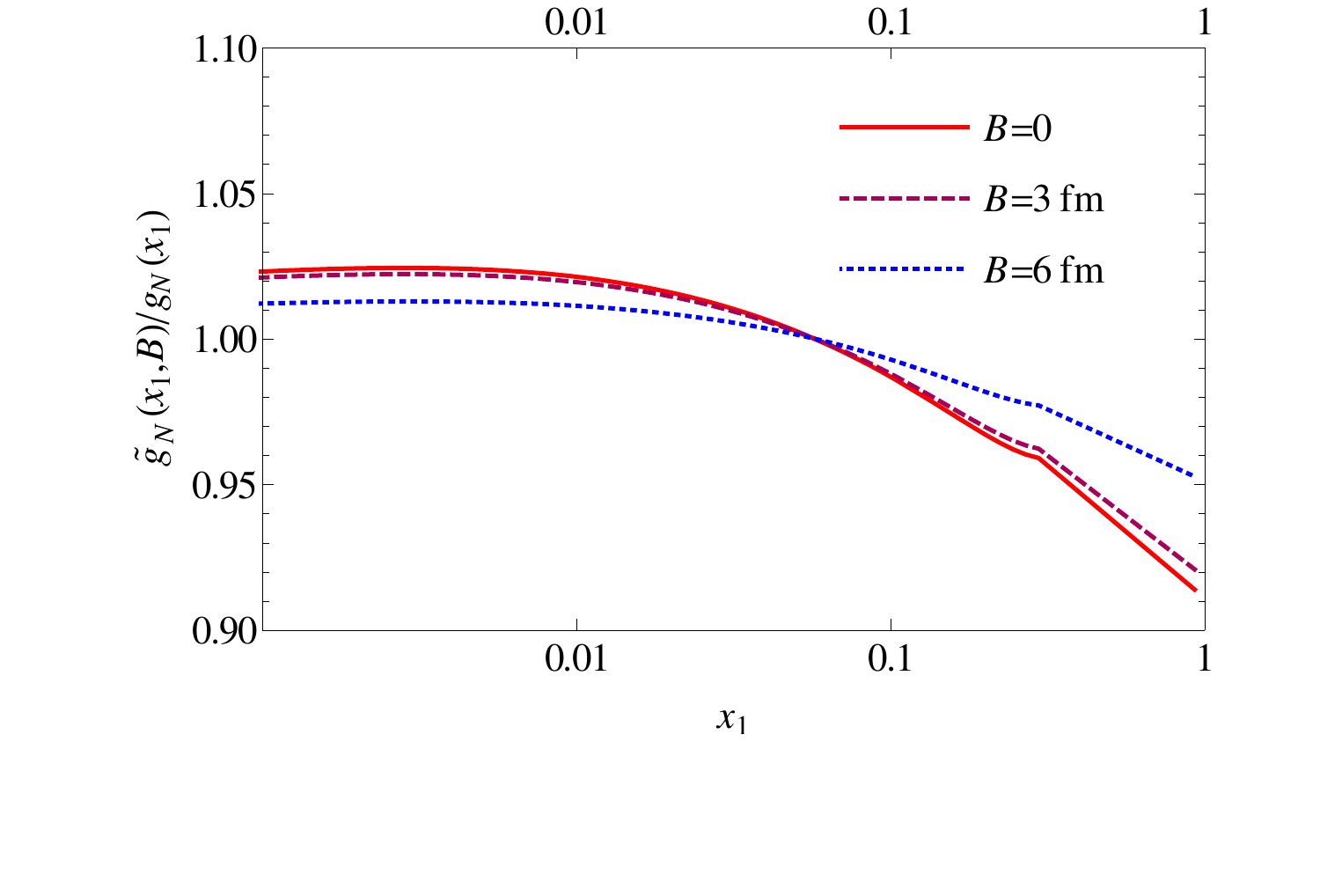}}}
  \vspace{-1cm}
\caption{\label{fig:dglap} (Color online) Ratio of the projectile gluon distributions in $p$-$Pb$ to $pp$ collisions at $\sqrt{s}=5000\GeV$  vs $x_1$ and $B$. 
The projectile gluon distribution, $\tilde g_N(x_1,B$) in $pA$ collisions is DGLAP evolved from the initial scale $4m_c^2$ to $4m_c^2+Q_s^2$, generated by the impact-parameter dependent saturation momentum $Q_s(B)$ .
}
 \end{figure}

With a larger  scale the process resolves more partons in the incoming proton. Thus,  via the effect of broadening the nuclear target activates higher Fock states in the incoming proton. The result is qualitatively similar to what we observed above, namely, parton density  will be enhanced at small $x_1$, but suppressed at $x_1\to1$. Such a nuclear modification of the gluon density in the incoming proton
can be performed by evolving the projectile proton PDFs with
DGLAP equations 
from the scale $Q^2=4m_c^2$ to $Q^2+Q_s^2$. Then the gluon PDF in the proton should be replaced $g_N(x_1)\Rightarrow \tilde g_N(x_1,B)$
in the numerators of $R^{(1N})$ and $R^{(2N})$.
Some examples of modifications, $\tilde g_N(x_1,B)/g_N(x_1)$ are shown in  Fig.~\ref{fig:dglap} for $pPb$ collisions at
$\sqrt{s}=5\TeV$ vs $x_1$ and impact parameter $B$.

This modification of the $x_1$ dependence of the projectile gluon distribution can be treated as an effective energy loss, leading to nuclear suppression of heavy quark production at forward rapidities (large $x_1$). The results presented in Fig.~\ref{fig:dglap} show that
the effect is extremely weak, only a few percent suppression at very forward rapidities.
The reason for this weakness can be easily traced in Fig.~1 of Ref.~\cite{boosting}.
One can see that the effect of induced energy loss is controlled by the relation between the scale of the process, $Q^2$, and the saturation scale $Q_s^2$. The effect may be strong if $Q_s^2\gg Q^2$, but becomes vanishingly small at $Q^2\gg Q_s^2$. 

Intuitively, this is pretty clear. It can be interpreted as a vacuum dead-cone effect \cite{hf-jets}, namely a parton originating from a hard process at scale $Q^2$ is lacking gluon field with small transverse momenta $k_T^2<Q^2$. Gluon bremsstrahlung and medium induced energy loss
of such a parton are significantly reduced compared to a nearly on-mass-shell parton.
This is what we see in the above Fig.~\ref{fig:dglap}, where the characteristic scale of the process, $Q^2\approx10\GeV^2$, exceeds considerably the saturation scale.

Reduction of induced energy loss by a large genuine scale $Q^2$ of the process
can be also interpreted in terms of the Landau-Pomeranchuk effect, which says that on a long length scale $l\gg R_A$, the radiation spectrum depends on the total accumulated kick
acquired by the charge, rather than on the details of several kicks occurring on a short length scale (the nuclear radius $R_A$). The radiation spectrum $dk_T^2/k_T^2$ leads to a logarithmic scale dependence of the radiated energy. The induced energy loss is given by a difference between energies radiated in the processes with the effective scales $Q^2+Q_s^2$ (in pA) and $Q^2$ (in pp). Thus, the induced energy loss exposes the following scale dependence,
\beq
\Delta E_{ind} \propto \ln\left(1+\frac{Q_s^2}{Q^2}\right)\approx \frac{Q_s^2}{Q^2},
\label{1310}
\eeq
if $Q^2\gg Q_s^2$, i.e. it turns out to be suppressed. This effect is of course included in the DGLAP analysis, whose results are presented in Fig.~\ref{fig:dglap}.

Notice that the suppressing effect of a large scale of the process was missed in the calculations 
 \cite{ap} of induced energy loss in charmonium production. As a result, the magnitude
 of energy loss was grossly overestimated compared with the DGLAP analysis.

\subsection{Numerical results for \boldmath$J/\psi$}

\begin{figure}[htb]
\centerline{
  \scalebox{0.65}{\includegraphics{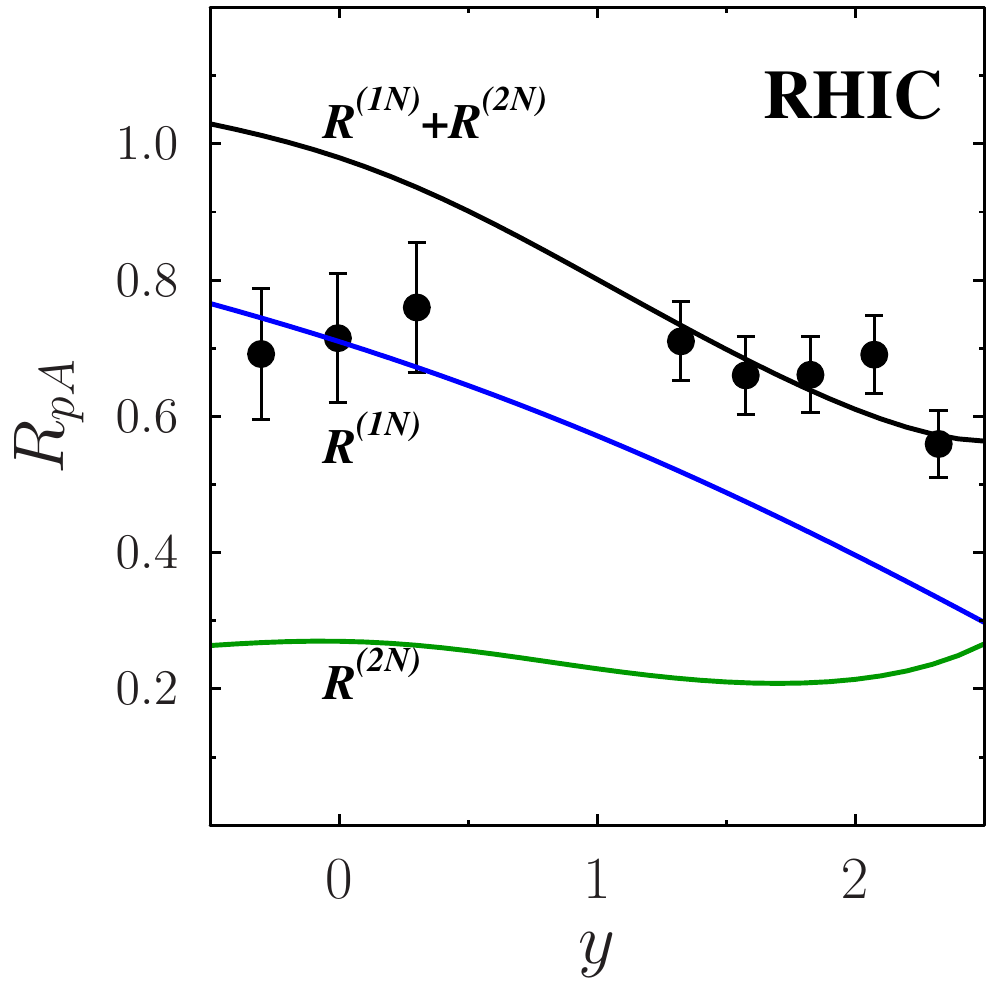}}}
\caption{\label{fig:rhic-full} (Color online) Ratio of $pAu$ to $pp$ cross sections of $J/\psi$ production at $\sqrt{s}=200\GeV$. The curves from bottom to top  present numerical results for the terms in Eq.~(\ref{440}) $R^{(2N)}$, $R^{(1N)}$, and their sum respectively. Gluon shadowing and nonperturbative and perturbative energy loss effects are included (see text). 
The data points are from \cite{phenix-psi}.
}
 \end{figure}

\begin{figure}[htb]
\centerline{
  \scalebox{0.65}{\includegraphics{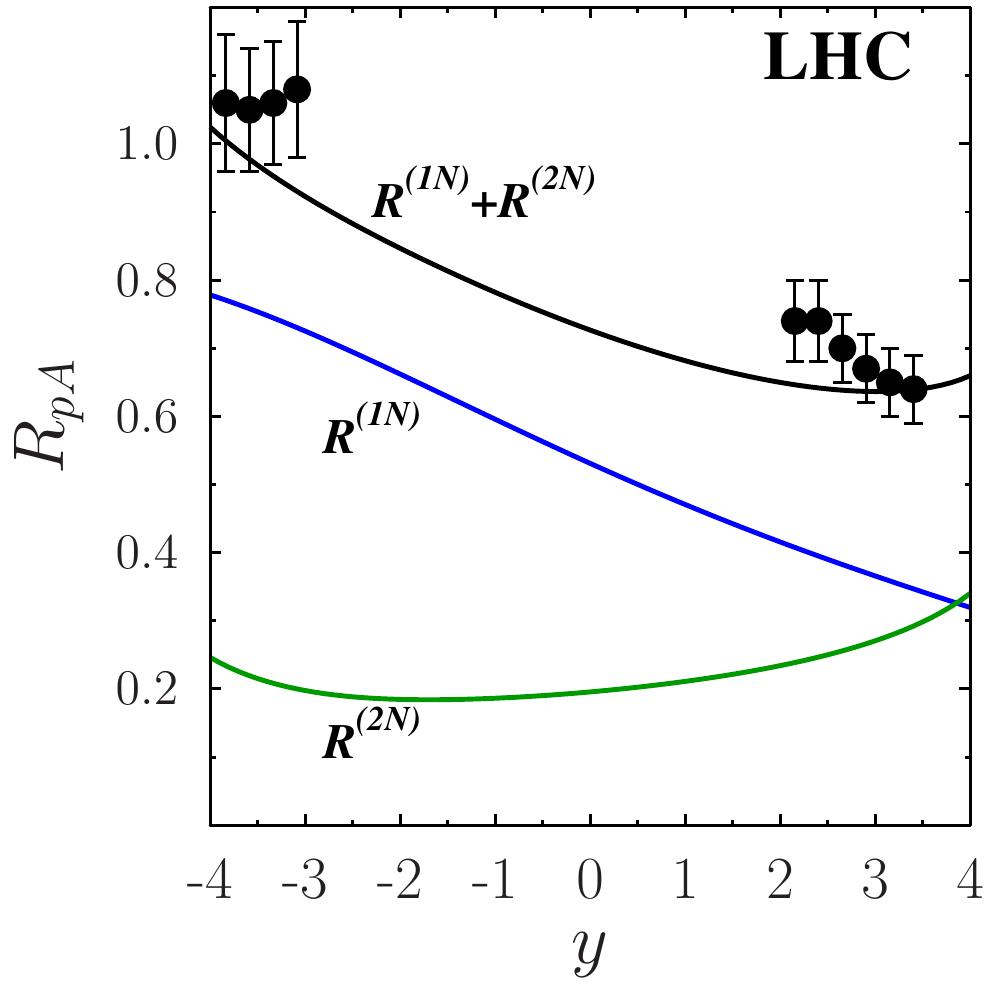}}}
\caption{\label{fig:lhc-full} (Color online) The same as in Fig.~\ref{fig:rhic-full}, but for $p$-$Pb$ collisions at $\sqrt{s}=5000\GeV$. Data points are from \cite{alice1,alice2}
}
 \end{figure}

Now we are in a position to finalize the calculations of nuclear effects in $J/\psi$ production.
The effects of energy loss, or modification of the projectile gluon distributions, have been already incorporated into our previous results corrected for gluon shadowing, as was plotted by the dashed curves in Fig.~\ref{fig:rhic-glue} and Fig.~\ref{fig:lhc-glue}. The final results are compared with available data at $\sqrt{s}=200\GeV$ in Fig.~\ref{fig:rhic-full} and at $\sqrt{s}=5000\GeV$ Fig.~\ref{fig:lhc-full}. 
As was anticipated, the energy loss effects are strongest at the energies of RHIC.
A substantial modification of nuclear effects due to energy loss has been already observed for other hard processes in \cite{high-pt,e-conserv}.
Our results seem to agree reasonably well with data, especially taking into account
the large uncertainties in the $pp$ reference, affecting the term $R^{(2N)}$ in (\ref{440}).

In view of the forthcoming LHC measurements of $pA$ collisions at $\sqrt{s}=8000\GeV$, we notice that our predictions are hardly different from those presented in Fig.~\ref{fig:lhc-full}
for $\sqrt{s}=5000\GeV$.

\subsection{Nuclear modification of the \boldmath$p_T$-distribution}

Multiple interactions of the projectile partons in the nucleus are known to lead to broadening of the transverse momentum, the phenomenon also called saturation or color glass condensate.
It can be effectively evaluated within the dipole phenomenology \cite{jkt}, well adjusted to HERA data on small-$x$ DIS. The value of broadening at impact parameter $B$ is given by
Eq.~(\ref{1300}) derived in \cite{jkt}. 

Nuclear broadening of the $p_T$-distribution naturally leads to a ratio $R_{pA}(p_T)$, rising with $p_T$, the effect, usually named after Cronin. The $p_T$ dependence of the 
$J/\psi$ production cross section in $pp$, $pA$ and $AA$ collisions is well described by the form, $d\sigma/dp_T^2\propto (1+p_T^2/6\la p_T^2\ra)^6$ \cite{e789,phenix-pt,alice2}. 
Therefore, making a shift of $\la p_T^2\ra$ for $pA$ in comparison with $pp$ collisions,
one arrives at a $p_T$-dependent nuclear ratio \cite{kps-psiAA},
\beq
R_{pA}(p_T)=R_{pA}\,{1\over\xi}
\left(\frac{1+p_T^2/6\la p_T^2\ra}
{1+p_T^2/6\xi \la p_T^2\ra}
\right)^6,
\label{1316}
\eeq 
where $R_{pA}$ in the r.h.s. of (\ref{1316}) is  the ratio of the $p_T$-integrated cross sections (as was calculated above); $\xi=1+\Delta_{pA}(x_2)/\la p_T^2\ra$; and $\Delta_{pA}(x_2)=\la p_T^2\ra_{pA}-\la p_T^2\ra_{pp}$ is nuclear broadening of 
charmonium transverse momentum.

\begin{figure}[htb]
\centerline{
  \scalebox{0.6}{\includegraphics{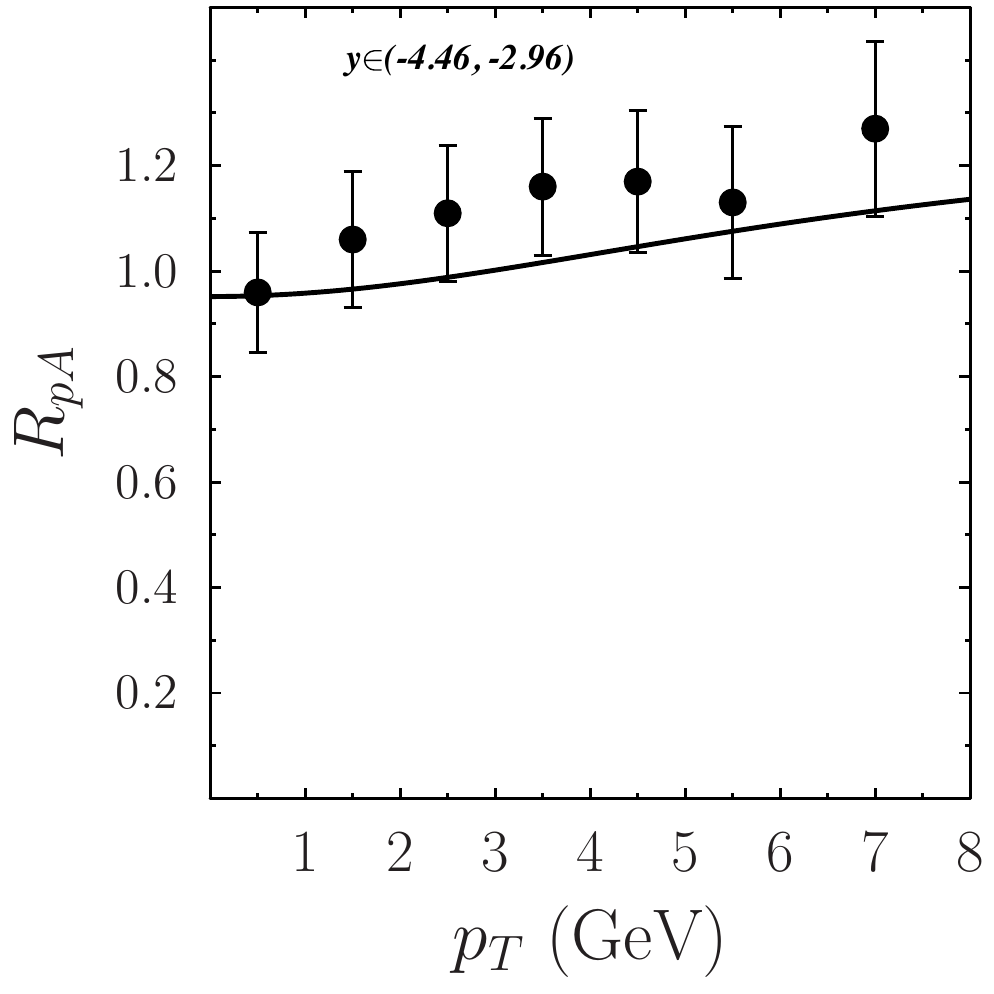}}}
\caption{\label{fig:R-pt1} The $p_T$-dependent ratio of the differential cross sections of inclusive (but direct) $J/\psi$ production in $pA$ and $pp$ collisions,
at $\sqrt{s}=5.02\TeV$ and $y\in(-4.46,-2.96)$.  Data points are from \cite{alice2}.
}
 \end{figure}

\begin{figure}[htb]
\centerline{
  \scalebox{0.6}{\includegraphics{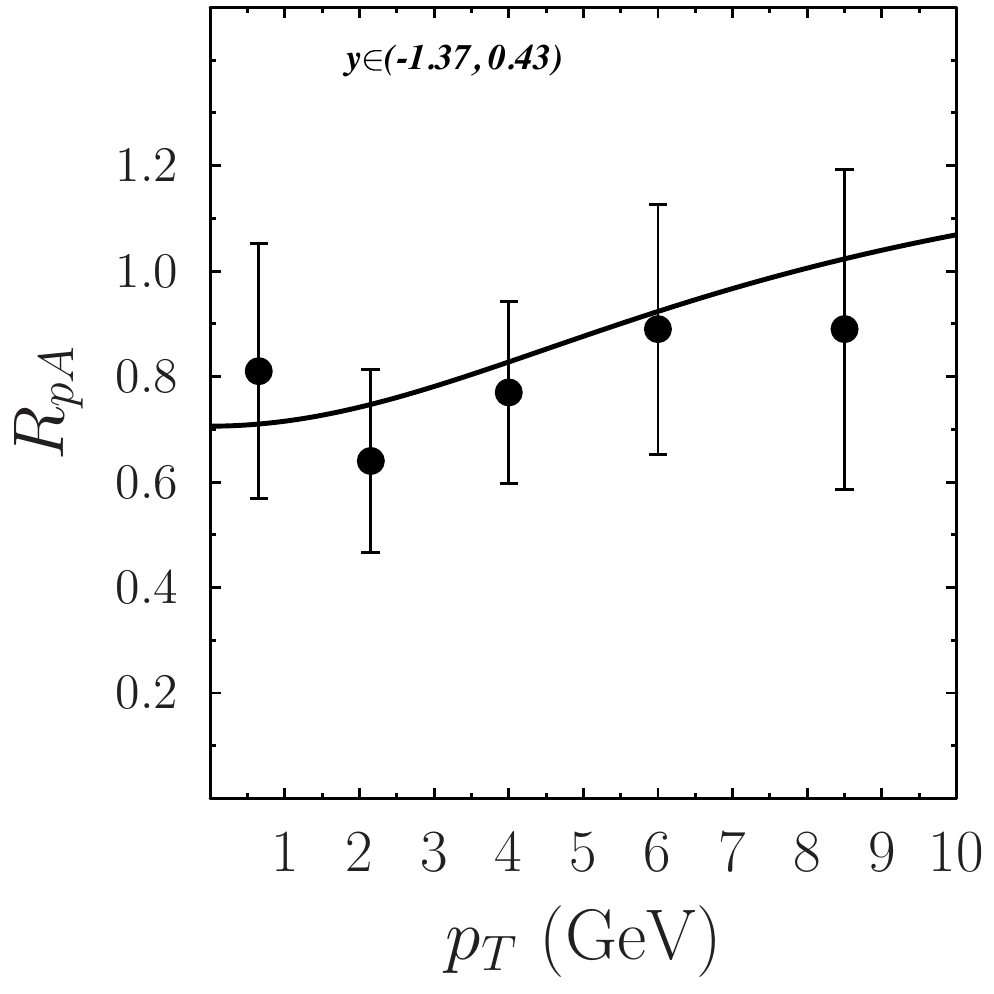}}}
\caption{\label{fig:R-pt2} The same as in Fig.~\ref{fig:R-pt1}, but for $y\in(-1.37,-0.43)$.
}
 \end{figure}

\begin{figure}[htb]
\centerline{
  \scalebox{0.6}{\includegraphics{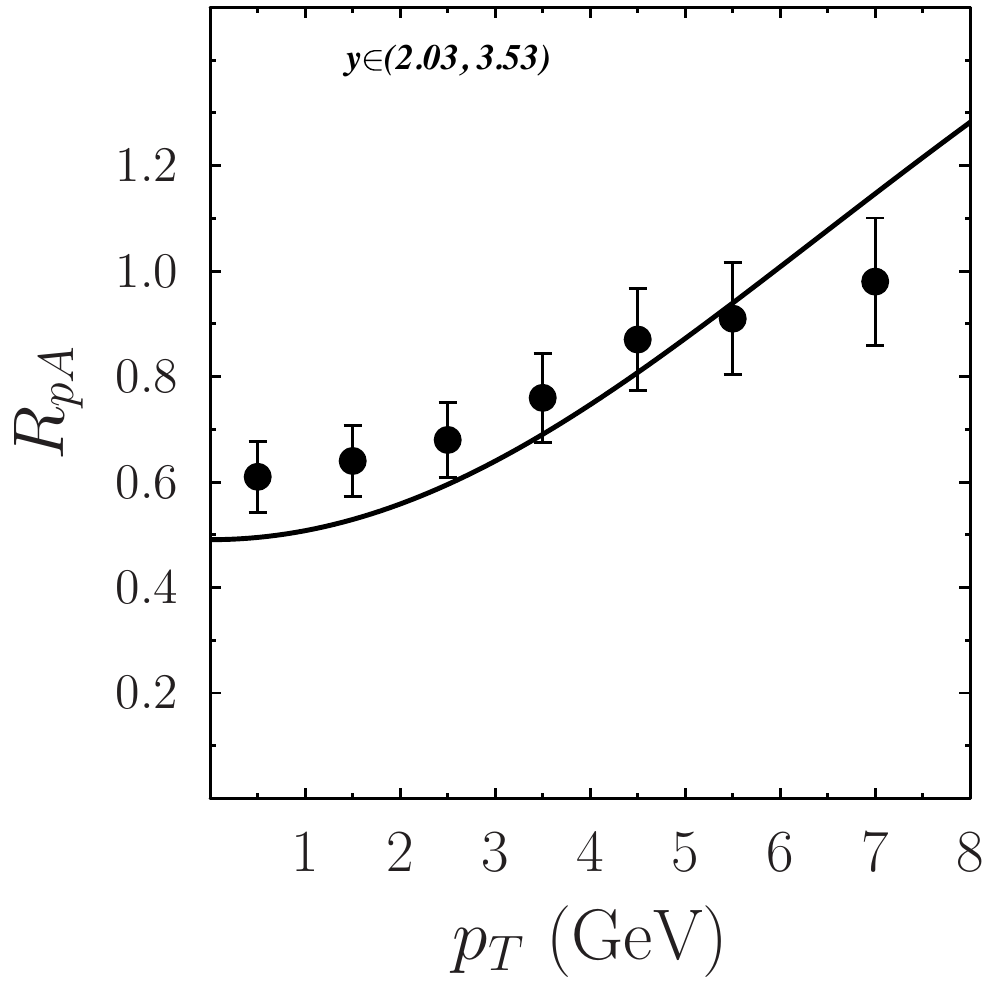}}}
\caption{\label{fig:R-pt3} The same as in Fig.~\ref{fig:R-pt1}, but for $y\in(2.03,3.53)$.
}
 \end{figure}

The magnitude of broadening was evaluated in \cite{broadening}. 
At $\sqrt{s}=5.02\TeV$ and the rapidity intervals of interest, $y\in(-4.46,-2.96)$,
$y\in(-1.37,-0.43)$ and $y\in(2.03,3.53)$ the broadening magnitudes, averaged over impact parameters, are 
$0.35$, $0.73$ and $2.27\GeV^2$ respectively.
The $p_T$-dependent $R_{pA}(p_T)$, given by Eq.~(\ref{1316}), calculated with these values and $\la p_T^2\ra=7\GeV^2$ \cite{alice2} 
are compared with data in figures~\ref{fig:R-pt1} - \ref{fig:R-pt3}, demonstrating good agreement.


\section{Production of \boldmath$\psi(2s)$}
\label{2s}

The first radial excitation $\psi(2S)$ has the mean radius squared about twice as large as that of $J/\psi$ \cite{buchmueller,cornell1,cornell2},
and therefore comparison of nuclear effects for these two charmonium states offers a sensitive test of the production dynamics. Expectations are usually based on either of two popular ideas,  {\sl both incorrect}: 

(i) The effect of color transparency makes the nuclear medium more transparent for
smaller size state, $J/\psi$, which is expected to be considerably less  suppressed than $\psi(2S)$. However, experiments at the SPS \cite{na3} and Fermilab \cite{e866} found similar magnitudes of nuclear suppression for the two charmonium states.

(ii) At first glance, the observed similarity of nuclear effects can be understood in line with the hierarchy of
characteristic length scales discussed in Sect.~\ref{time-scales}. Indeed, at high energies the formation length Eq.~(\ref{1010}) substantially exceeds the nuclear dimension, so a perturbatively small $\bar cc$ dipole, rather than a formed charmonium of much larger larger size, propagates through the nucleus. Then one expects the dipole to evolve into either $J/\psi$ or $\psi(2S)$ outside of the nucleus, after experiencing an universal nuclear attenuation on the early perturbative stage. Naively, one might expect universal nuclear suppression for different charmonia.
However, the dynamics, controlling the nuclear effects is more involved.

The second proposal (ii) explains why the first one, (i), is incorrect.
Nonetheless, an universal nuclear attenuation of a $\bar cc$ dipole does not lead to an universal charmonium suppression, because the projection of the produced $\bar cc$ distribution function to the charmonium wave function depends on the latter. In particular,
spectacular  effects are expected  for production of $\psi(2S)$,  related to the specific
shape of its wave function, which has a node and changes sign as function of the $\bar cc$ separation.

Unusual features of $\psi(2S)$ production were revealed in photoproduction of charmonia \cite{kz91}, the process of a similar, although simpler  dynamics compared with hadro-production.
It was found that in spite of its large size, the $\psi(2S)$ produced in nuclei may be less suppressed compared with $J/\psi$, sometimes even enhanced. This can be interpreted either in terms of the multi-channel generalised  Glauber model \cite{hufner96}, or
within the dipole description as a result of the specific nodal structure of the $\psi(2S)$ 
wave function \cite{kz91,knnz93,hikt2}. The $\bar cc$ distribution function, to be projected 
to the charmonium wave function, has a rather wide $r$-distribution, which peaks at $r\sim 2/m_c$ \cite{nontrivial,knnz93}, close to the node position in the $\psi(2S)$ wave function. Therefore, a part of the overlap integral extends beyond the node and contributes with a negative sign, causing a significant compensation between dipole separations smaller and larger than the node position.  This cancellation contributes to the observed suppression of $\psi(2S)$ production \cite{knnz93} in $pp$ collisions. A nuclear target serves as a color filter, which removes the large-size $\bar cc$ dipoles, and therefore the mean size of the $\bar cc$ wave packet is reduced and the overlap with the $\psi(2S)$ wave function increases.

\begin{figure}[htb]
\centerline{
  \scalebox{0.65}{\includegraphics{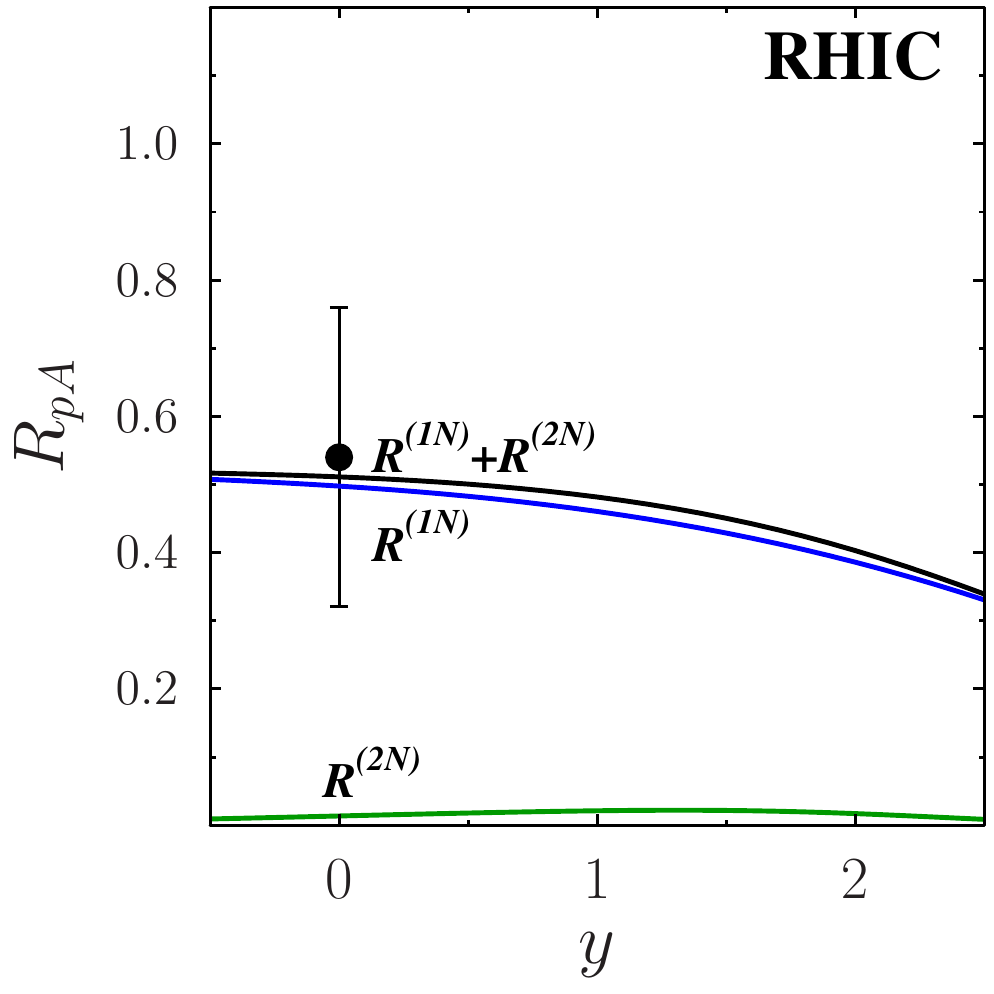}}}
\caption{\label{fig:psi2s-rhic} (Color online) Ratio of $pAu$ to $pp$ cross sections of $\psi(2S)$ production at $\sqrt{s}=200\GeV$. The curves from bottom to top present numerical results for the terms in Eq.~(\ref{440}) $R^{(2N)}$, $R^{(1N)}$, and their sum respectively. Gluon shadowing, as well as the nonperturbative and perturbative energy loss effects are included (see text). 
The data point is from \cite{phenix-2s}.
}
 \end{figure}

\begin{figure}[htb]
\centerline{
  \scalebox{0.65}{\includegraphics{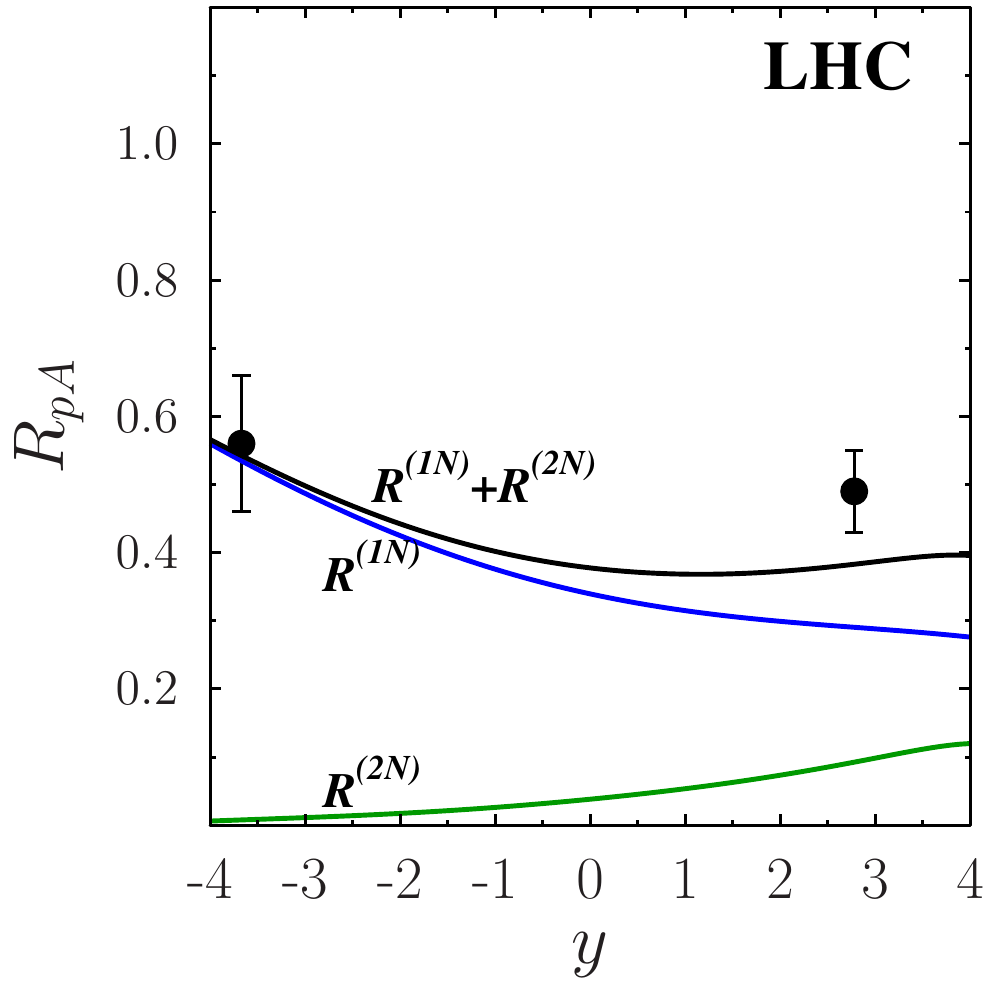}}}
\caption{\label{fig:psi2s-lhc} (Color online) The same as in Fig.~\ref{fig:psi2s-rhic}, but for $p$-$Pb$ collisions at $\sqrt{s}=5000\GeV$. Data points are from \cite{alice-2s}
}
 \end{figure}

The results of calculations of the nuclear ratio $R_{pA}$ for $\psi(2S)$ are compared with
available data at RHIC and LHC in Figs.~\ref{fig:psi2s-rhic} and  \ref{fig:psi2s-lhc}, respectively.
The double scattering term $R^{(2N)}$ turns out to be very small for $\psi(2S)$ at the energies of RHIC, but rises to a sizeable corrections at higher energies.

\begin{figure}[htb]
\centerline{
  \scalebox{0.65}{\includegraphics{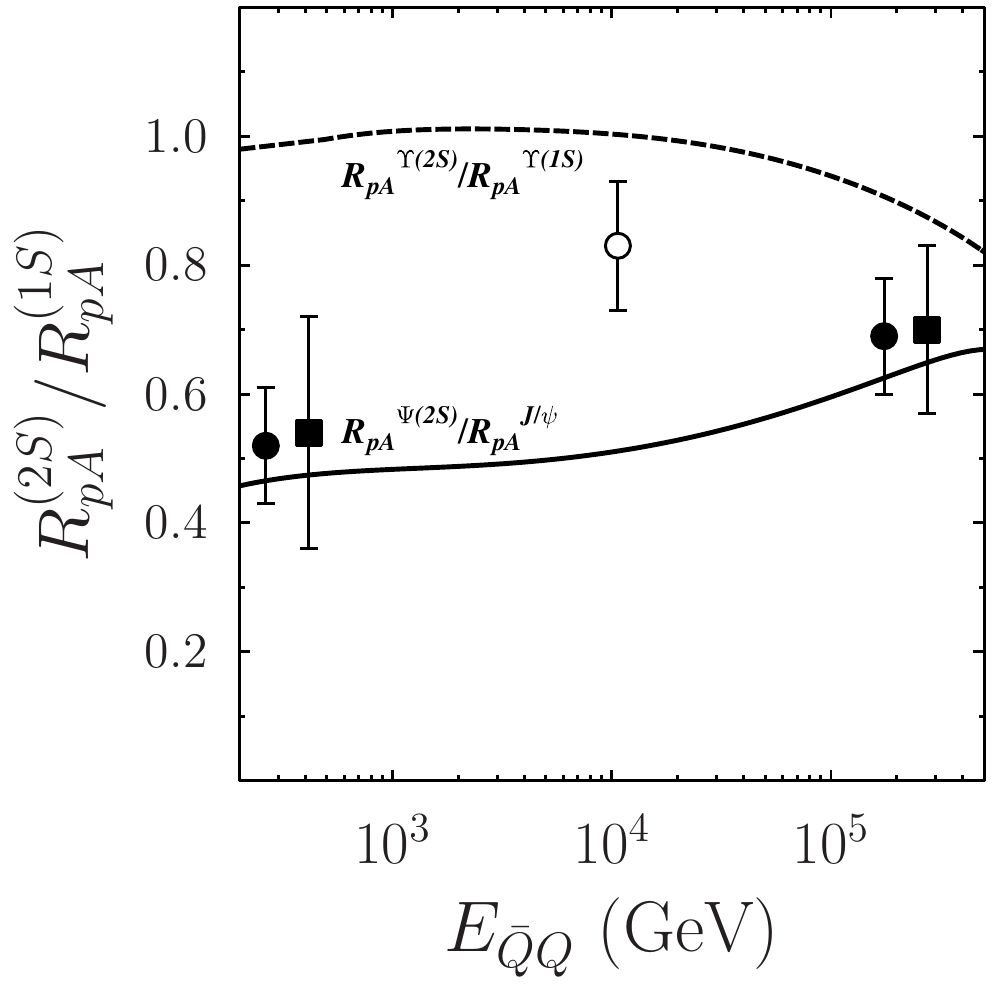}}}
\caption{\label{fig:double-ratio} The double ratio $R_{pA}^{(2S)}/R_{pA}^{(1S)}$ as function of quarkonium energy in the nuclear rest frame, $E_{\bar QQ}=M_{\bar QQ}^2/2x_2m_N$. Solid and dashed curves show the results of calculations for charmonium and bottomium, respectively. Green full circles and squares show the results of  respectively ALICE \cite{alice-2s} and LHCb \cite{lhcb-double} measurements of $R_{pA}^{\psi(2S)}/R_{pA}^{J/\psi}$ at $\sqrt{s}=5.02\TeV$.
The blue empty circle shows the CMS result \cite{cms-upsilon-double} for $\Upsilon(2S)/\Upsilon(1S)$
$pPb$ to $pp$ double ratio at $\sqrt{s}=5.02\TeV$.
}
 \end{figure}

Again, we can conclude that our calculations do not contradict data, which has rather large errors.
However, our results for the double ratio $R_{pA}^{\psi(2S)}/R_{pA}^{J/\psi}$, plotted in Fig.~\ref{fig:double-ratio}, show rather small values slowly rising with energy. These results contradict the precise data of E866 experiment \cite{e866}, which show that at small $x_F$ the double ratio is about $R_{pA}^{\psi(2S)}/R_{pA}^{J/\psi}=0.9$, with a small error.

The nuclear  effects observed for the production of the first radial excitation $\psi(2s)$ demonstrate 
suppression, similar to $J/\psi$, in the energy range of fixed target experiments \cite{na3,e866}. However, in the energy range of RHIC-LHC, a stronger suppression of $\psi(2s)$ relative to $J/\psi$ was observed \cite{phenix-2s,alice-2s}. 

\section{Upsilon production}
\label{upsilon}

\begin{figure}[htb]
\centerline{
  \scalebox{0.65}{\includegraphics{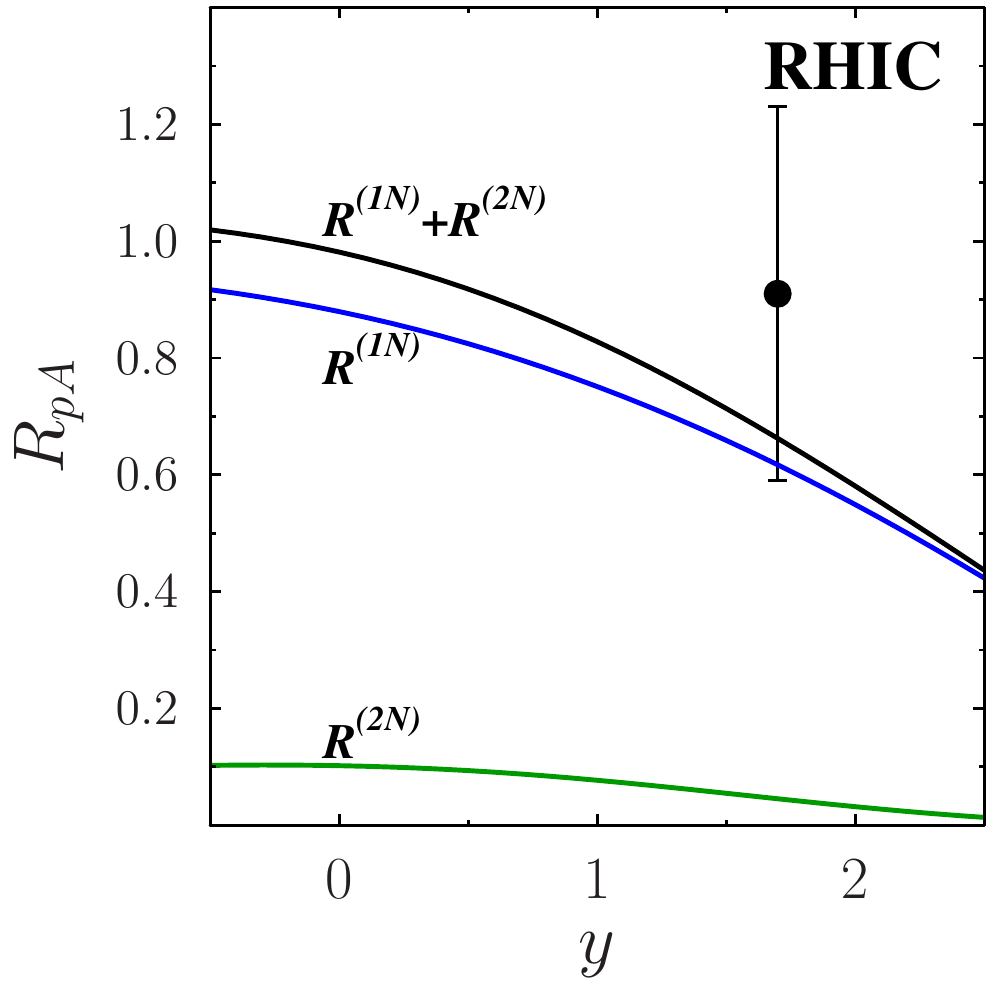}}}
\caption{\label{fig:upsilon-rhic} (Color online) The same as in Fig.~\ref{fig:rhic-full}, but for $\Upsilon$ production in $p$-$Au$ collisions at RHIC at $\sqrt{s}=200\GeV$.
The data point is from \cite{phenix-upsilon}.
}
 \end{figure}

\begin{figure}[htb]
\centerline{
  \scalebox{0.65}{\includegraphics{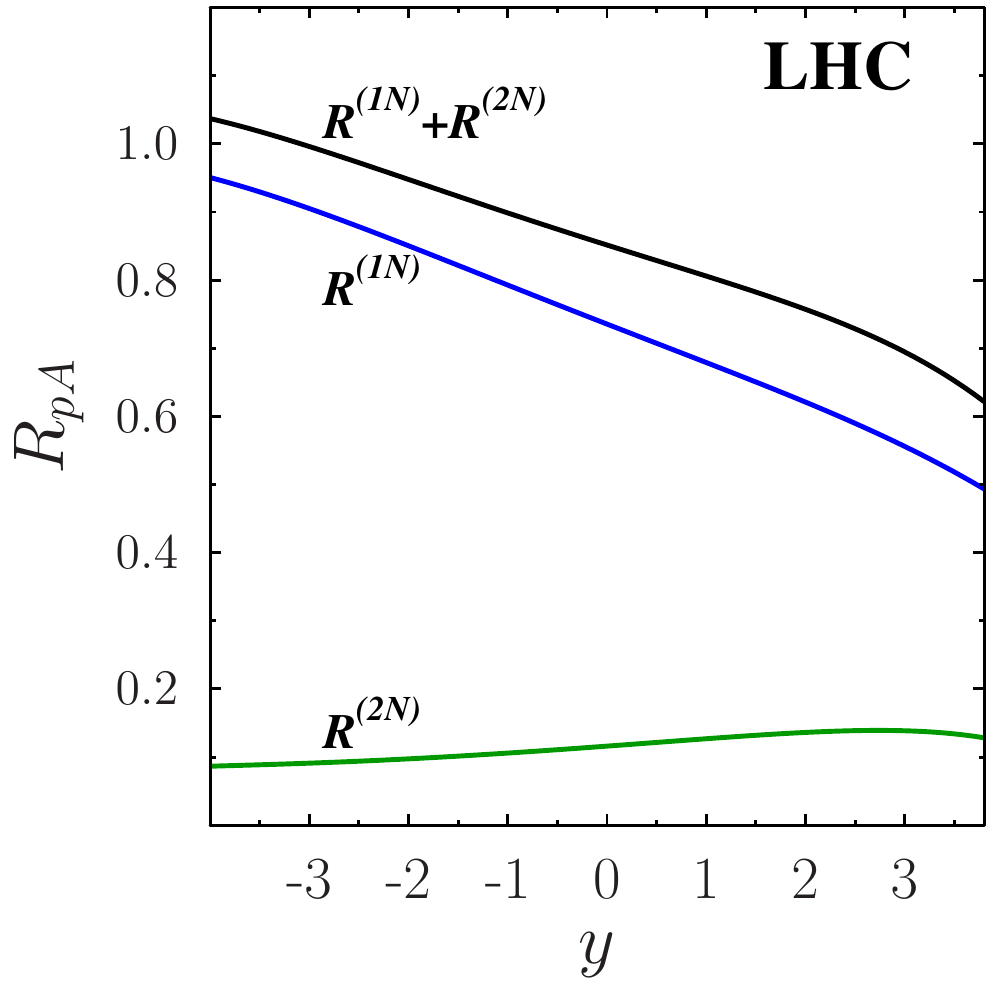}}}
\caption{\label{fig:upsilon-lhc} (Color online) The same as in Fig.~\ref{fig:upsilon-rhic},
but at $\sqrt{s}=5000\GeV$.
}
 \end{figure}

The developed dipole description of charmonium production in $pA$ collisions can be naturally extended for bottomium production, replacing the charm quark mass by $m_{b}=4.5\GeV$.
In the Figures~\ref{fig:upsilon-rhic} and \ref{fig:upsilon-lhc} we present the results at the energies of RHIC and LHC respectively. The term $R^{(1)}$ closely reproduces
the earlier calculations in \cite{nontrivial}, except for the added energy loss effect, which affects the results for RHIC, but not for LHC.

Due to larger $b$-quark mass and smaller dipole sizes, the two-nucleon term $R^{(2N)}$ in (\ref{440})  is relatively smaller compared with $J/\psi$ production, as one can see in Figs.~\ref{fig:upsilon-rhic} and \ref{fig:upsilon-lhc}. As for charmonium, we calculate the $pp$ reference cross section used in the denominator of $R^{(2N)}$, within the CSM, and adjust its normalization to data.

The only available data point \cite{phenix-upsilon}, plotted in fig.~\ref{fig:upsilon-rhic}
has too large error bar to be considered as a support for our calculations.

We also performed calculations for the double ratio $R_{pA}(\Upsilon(2S))/R_{pA}(\Upsilon(1S))$, and plotted it as function of $\bar bb$ energy in Fig.~\ref{fig:double-ratio}.
This ratio was measured with a good precision in the CMS experiment at $\sqrt{s}=5000GeV$
and $|y|<1.93$ \cite{cms-upsilon-double}. This point, plotted in Fig.~\ref{fig:double-ratio}
at energy $E_{\bar bb}=e^y M_{\Upsilon}\sqrt{s}/2m_N$, agree well with our parameter-free calculations.

\section{Summary and conclusions}

The main objective of this work was to settle the problem of  the energy independence of nuclear effects for $J/\psi$ production, observed in $pA$ collisions. This independence of energy is in striking contradiction with the steep energy dependence of the dipole cross section observed at HERA , which controls the nuclear effects. We revealed a novel mechanism enhancing charmonium production at high energies, which comes from the next order of the opacity expansion.

Crucial for the results was the choice of mechanism dominating the production of heavy flavor vector mesons in pp collisions. We favored the color-singlet model (CSM), which
can dominate the small-$pT$ quarkonium production we are interested in. 
We developed a color-dipole formulation of CSM, which is crucial for the calculation of nuclear effects. 

The second order term in the opacity expansion for the production cross section is dominated by a different mechanism, a double color-exchange interaction of the projectile heavy $\bar QQ$ dipole. Its contribution helps to reach agreement with data for the nuclear suppression of $J/\psi$ production both at the energies of RHIC and LHC.

Other nuclear effects, gluon shadowing and energy loss, have also been included. Gluon shadowing corrections are found to be important at the energies of LHC, but very small at RHIC. On the contrary, energy loss effects substantially suppress quarkonium production rates
at forward rapidities at RHIC, while have no influence at the energies of LHC.
The main contribution comes from the nonperturbative mechanism of energy loss, related to the energy sharing problem at forward rapidities. The perturbative energy loss generated 
by $p_T$ broadening was found to be suppressed by the smallness of the saturated momentum relative the scale of the process. This suppression was missed in the previous calculations of the energy loss effect, which has been grossly overestimated.

Although we restricted these calculations with the $p_T$-integrated cross sections, the $p_T$-dependent ratio $R_{pA}(p_T$ was also evaluated, based on the known empirical shape of the $p_T$-distribution and the value of broadening, calculated in a parameter-free way (although not free of assumptions) within the dipole phenomenology. The results, obtained for several rapidity intervals, well
agree with ALICE data.

Production of radial excitations, vector quarkonia in the $2S$ state, has always attracted interest, related to the nodal structure of the wave function. Differently from photoproduction,
where $2S$ states are enhanced compared with the ground state, in hadroproduction we found a strong nuclear suppression of the $\psi(2S)$ to $J/\psi$ ratio, in good agreement with data.
At the same time, for bottomia, the $2S$ to $1S$ ratio is nearly unaffected by the nuclear effects, what could be anticipated, because the $\bar bb$ dipoles are much smaller compared with $\bar cc$, so the convolution with the $\Upsilon$ wave function is less important.

\begin{acknowledgments}
This work was supported in part
by Fondecyt (Chile) grants 1170319, 1140842, 1140390 and 1140377,
by Proyecto Basal FB 0821 (Chile),
and by CONICYT grant  PIA ACT1406 (Chile) .
Powered@NLHPC: This research
was partially supported by the supercomputing infrastructure of the
NLHPC (ECM-02). Also, we thank Yuri Ivanov for technical support of
the USM HPC cluster where a part of evaluations has been done.

\end{acknowledgments}

 \def\appendix{\par
 \setcounter{section}{0}
\setcounter{subsection}{0}
 \def\thesection{Appendix \Alph{section}}
\def\thesubsection{\Alph{section}.\arabic{subsection}}
\def\theequation{\Alph{section}.\arabic{equation}}
\setcounter{equation}{0}}

 \appendix
 
\section{Multiple color-exchange interactions of a high-energy dipole}
\label{alesha}
\setcounter{equation}{0}

At sufficiently high energy, when the length scales discussed in Sect.~\ref{time-scales} considerably exceed the nuclear dimensions, one can treat the transverse size of such a dipole as "frozen" by Lorentz time dilation during propagation through the nucleus. 
The kinematic constraints for this regime can be found in Sect.~\ref{frozen}.
This is a perturbative stage of interaction, so the one-gluon approximation for dipole-nucleon interaction is justified. However, multigluon-exchange interactions with different nucleons
are enhances by powers of $A^{1/3}$ and cannot be neglected.

\subsection{Evolution of the \boldmath$\bar cc$ density matrix}

\begin{figure}[htb]
\centerline{
  \scalebox{0.17}{\includegraphics{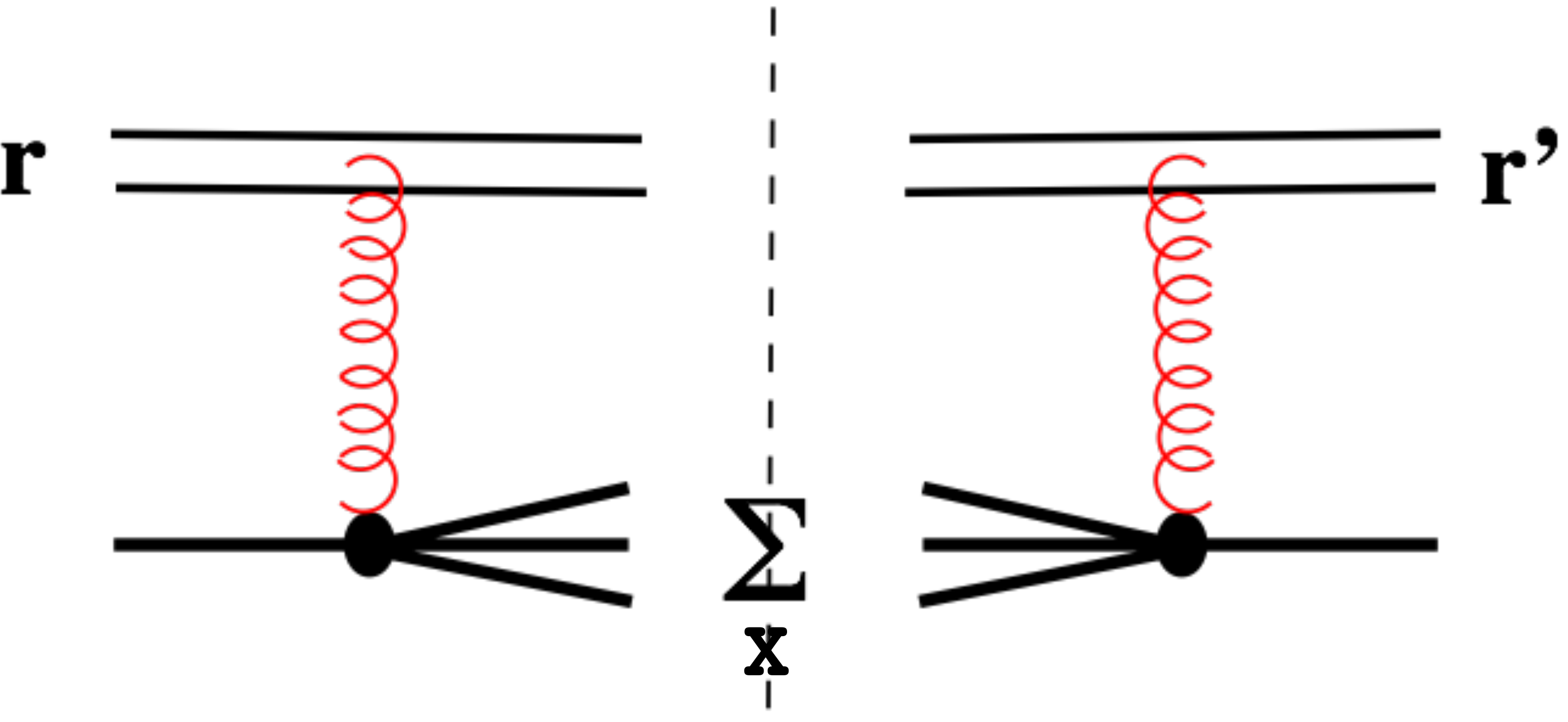}}}
\caption{\label{fig:dens-matr} (Color online) Inelastic dipole-nucleon amplitude, squared and summed over final nucleon debris.}
 \end{figure}

\begin{figure}[htb]
\centerline{
  \scalebox{0.4}{\includegraphics{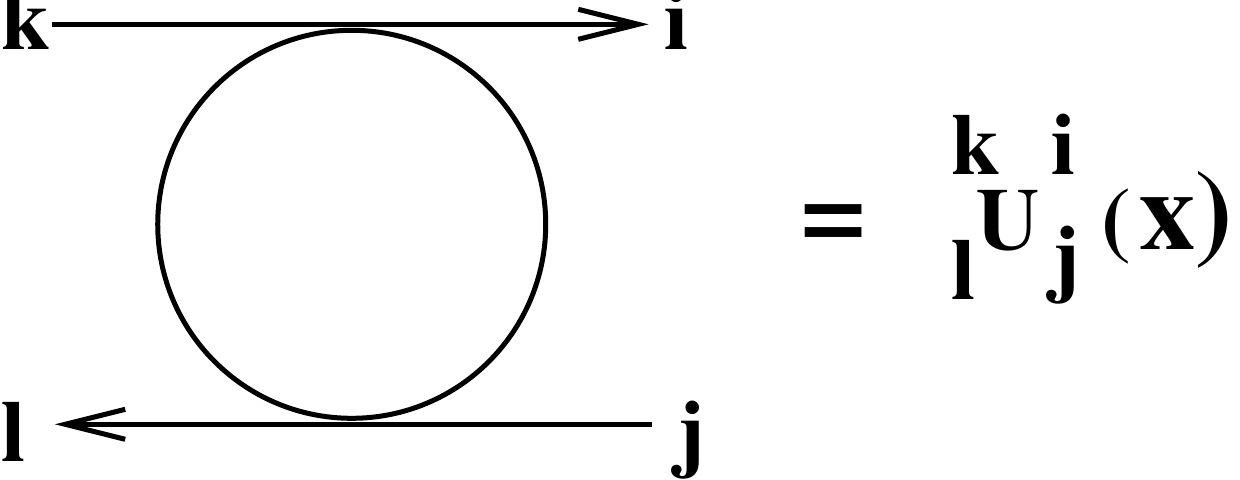}}}
\caption{\label{fig:dm} (Color online) Graphical representation of the density matrix, describing color states of the interacting dipole.}
 \end{figure}

Multiple soft color-exchange interactions of with the bound nucleons keep the dipole transverse separation $\vec r$ unchanged, but destroy the target, 
\beq
\bar c^{\,i}c_j+N\to \bar c^{\,k}c_l+X,
\label{a120}
\eeq
as is illustrated in Fig.~\ref{fig:multiple}.
One cannot describe the dipole evolution in terms 
of the dipole-nucleus amplitude, because in the cross section
the final states of each color-exchange collision must be summed-up,
as is illustrated in Fig.~\ref{fig:dens-matr}.
Therefore, the dipole propagation in the medium is described in terms of density matrix
$\hbox{}^k_l U^i_j(x_1,x_2;x_1^\prime,x_2^\prime)$,
where $x_{1,2}$ and $x_{1,2}^\prime$ are the transverse coordinates of the quark and antiquark in the two conjugated amplitudes \cite{alesha,k-zam,kth-psi},
which presented graphically in Fig.~\ref{fig:dm}.
We will follow the evolution of the density matrix along the longitudinal
coordinate $z$, which measures the propagation of the system through
the nucleus.

Before the $\bar cc$ pair enters the nucleus, i.e. at $z\to -\infty$, it 
is in a pure colorless state, i.e.
\beqn
&&\left.\hbox{}^k_l U^i_j
(\vec x_1,\vec x_2;\vec x_1',\vec x_2';z)
\right|_{z\to-\infty}
\nonumber\\&=&
\Psi_{in}
(\vec x_1-\vec x_2)\Bigr|^i_j
\left.\Psi^{\dagger}_{in}(\vec x_1'-\vec x_2')\right|^k_l,
\label{a140}
\eeqn
where $\Psi_{in}(r)$ is the distribution function of $\bar cc$ in the incoming
beam, for instance a $\bar cc$ component of a projectile gluon.

At $z\to\infty$ the system leaves the nucleus and the density
matrix can be projected directly to the final state wave function, 
\beqn
&&\int \prod\limits_{n,m} d^2x_n d^2x^\prime_m\, 
\hbox{}^k_l U^i_j
(\vec x_1,\vec x_2;\vec x_1',\vec x_2';z)\Bigr|_{z\to\infty}
\nonumber\\ &\times&
\Psi_{f}
(\vec x_1-\vec x_2)\Bigr|^j_i
\left.\Psi^{\dagger}_{f}(\vec x_1'-\vec x_2')\right|^l_k,  
\label{a160}
\eeqn

Since for every interaction of the $\bar cc$ in the medium we sum up over the final states of nucleons, the density matrix is a
colourless object, i.e. it is invariant under simultaneous rotations
in all colour indices $i,j,k,l$. Therefore it can be conveniently
decomposed into the irreducible parts corresponding to singlet and
octet states of the pair,
\beq
\hbox{}^k_l U^i_j
(\vec r;\vec r^{\,\prime};z)=S(\vec r;\vec r^{\,\prime};z)\,P_S+
{1\over8}\,O(\vec r;\vec r^{\,\prime};z)\,P_O,
\label{240}
\eeq
where $z$ is longitudinal coordinate of the target nucleon; $\vec r=\vec x_1-\vec x_2$, $\vec r^{\,\prime}=\vec x_1^{\,\prime}-\vec x_2^{\,\prime}$. We assume here that the impact parameters of the centers of gravity of the dipoles in the two amplitudes coincide, which is correct if the dipole-nucleon interaction radius can be neglected compared with the nuclear radius.

$P_S$ and $P_O$ in (\ref{240}) are the singlet and octet projection operators,
\beqn
P_S&=&{1\over 3}\,\delta^i_j\delta^k_l;
\nonumber\\
P_O&=&\delta^i_l\delta^k_j-{1\over 3}\,\delta^i_j\delta^k_l
\label{a260}
\eeqn
such that
\beqn
{\rm Tr}\,P_S=1 ;
\ \ \ \ \ {\rm Tr}\,P_O=8.
\label{a280}
\eeqn

The elements $S(\vec r=\vec r^{\,\prime})$ and $O(\vec r=\vec r^{\,\prime})$ are the probabilities
to find the quark-antiquark pair in color singlet or octet states respectively. 

In the one-gluon-exchange model every interaction with a
nucleon results in the change of the density matrix 
$\hbox{}^k_l U^i_j$, represented schematically in fig.3.
\begin{widetext}
\begin{center}
\begin{figure}[htb]
\centerline{
  \scalebox{0.4}{\includegraphics{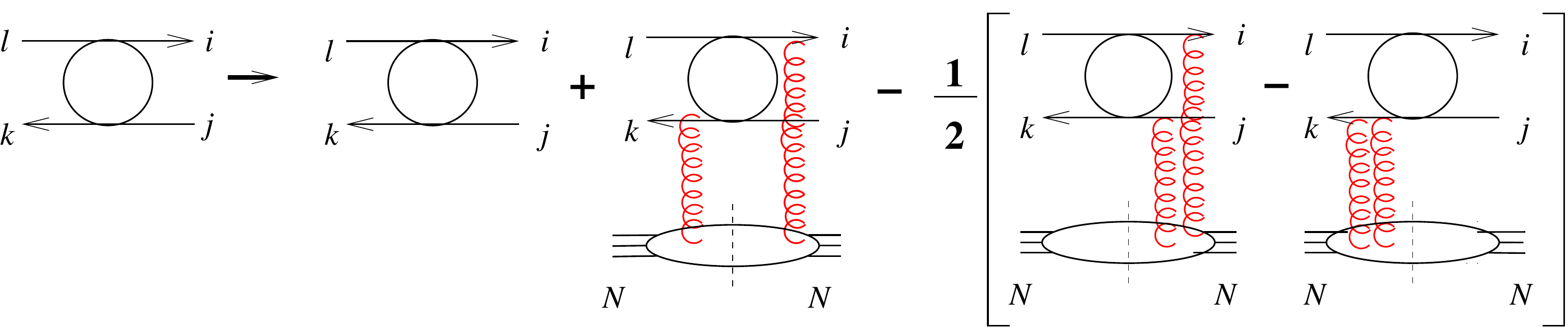}}}
\caption{\label{fig:evol} (Color online) Different unitarity cuts of the dipole-nucleon interaction cross section. The unitarity cuts are shown by dashed lines.}
 \end{figure}
\end{center}

Explicit calculation of the diagrams gives the following variations of the density matrices as function of $z$,
\beqn
\frac{d}{dz}S(\vec r,\vec r^{\,\prime};z)&=&
\biggl[-\Sigma_1(\vec r,\vec r^{\,\prime}) S(\vec r,\vec r^{\,\prime};z)
+ \Sigma_{tr}(\vec r,\vec r^{\,\prime}) O(\vec r,\vec r^{\,\prime}\biggr]n_A(b,z);
\label{a285}\\
\frac{d}{dz}O(\vec r,\vec r^{\,\prime};z)&=&
\bigl[8\Sigma_{tr}(\vec r,\vec r^{\,\prime})\,S(\vec r,\vec r^{\,\prime};z)
- \Sigma_8(\vec r,\vec r^{\,\prime})\,O(\vec r,\vec r^{\,\prime};z)\bigr]n_A(b,z),
\label{a300}
\eeqn
\end{widetext}

where
\beq
\Sigma_1(\vec r,\vec r^{\,\prime})=
{1\over2}\bigl[\sigma_{\bar qq}(r)+\sigma_{\bar qq}(r^\prime)\bigr];
\label{305}
\eeq
\beqn
\Sigma_{tr}(\vec r,\vec r^{\,\prime})=
{1\over8}\left[
\sigma_{\bar qq}\left(\frac{\vec r+\vec r^{\,\prime}}{2}\right)-
\sigma_{\bar qq}\left(\frac{\vec r-\vec r^{\,\prime}}{2}\right)\right];
\label{310}
\eeqn
\beqn
\Sigma_{8}(\vec r,\vec r^{\,\prime})&=&
{1\over8}\biggl[
4\sigma_{\bar qq}\left(\frac{\vec r+\vec r^{\,\prime}}{2}\right)
+14\sigma_{\bar qq}\left(\frac{\vec r-\vec r^{\,\prime}}{2}\right)
\nonumber\\  &-&
\sigma_{\bar qq}(r)-\sigma_{\bar qq}(r^\prime)
\biggr]
\label{a320}
\eeqn

If one is not interested in a particular spacial state of the outgoing
$\bar cc$ pair and regarding only its colour state (e.g., one does
not descriminate between different outgoing colourless
states like $J/\psi$, $\eta_c$, $\chi$, etc.), only the elements diagonal in the space variables $\vec x_1=\vec x_1'$ and
$\vec x_2=\vec x_2'$ of the density matrix are relevant.
 Then for
$S(\vec r;z)$ and $O(\vec r;z)$, which also implicitly depend on $b$,
one gets the following system of linear differential equations
\beq
{d\over dz}S(r;z)=\left[-S(r;z)+{1\over 8}
O(r;z)\right]n_A(b,z)\sigma_{\bar qq}(r)
\label{a330}
\eeq
Here $S(r;z)$ and $O(r;z)$ are interpreted
as the probabilities to find the $\bar cc$ pair in a color singlet or
octet states respectively. Since the total probability is conserved,
\beq
{d\over dz}\bigl[S(r;z)+O(r;z)\bigr]=0
\label{a335}
\eeq

Assuming that the initial state is a pure singlet with distribution function $S_{in}(r)$,
and solving Eqs.~(\ref{a330})-(\ref{a335}) one arrives at,
\beqn
S(r,z)&=&\left[{1\over 9}+{8\over 9}\,e^{-{9\over 8}\sigma_{\bar qq}
(r)T_A(b,z)}\right]S_{in}(r);
\nonumber\\
O(r,z)&=&\left[{8\over 9}-{8\over 9}\,e^{-{9\over 8}\sigma_{\bar qq}
(r)T_A(b,z)}\right]S_{in}(r).
\label{a340}
\eeqn
Correspondingly, for a color-octet initial state one gets,
\beqn
S(r,z)&=&\left[{1\over 9}-{1\over 9}\,e^{-{9\over 8}\sigma_{\bar qq}
(r)T_A(b,z)}\right]O_{in}(r);
\nonumber\\
O(r,z)&=&\left[{8\over 9}+{1\over 9}\,e^{-{9\over 8}\sigma_{\bar qq}
(r)T_A(b,z)}\right]O_{in}(r).
\label{a380}
\eeqn

We see that for large number of inelastic collisions of the $\bar cc$ dipole\footnote{One should not mix up this value with the number of collision usually used for normalization of hard reactions in $pA$ and $AA$ collisions. The latter is controlled by $\sigma_{in}^{NN}$,
rather than by the small $\bar cc$ dipole cross section.}, $n^{\bar cc}_{coll}=\sigma_{\bar cc}(r)T_A(b,z)\gg1$
the probability of production of color-singlet or octet states approach universal values, $1/9$ and $8/9$ respectively, independently of the color structure of the incoming $\bar cc$ pair.
This could be anticipated, since after multiple rotations in the color space both quark become completely unpolarized in color. All of the possible $9$ ($3\times3$) color states of the $\bar cc$ are produced with equal probabilities, and only one of them is a singlet, while the 8 others are octets.

\section{ \boldmath  $J/\psi$ production in $pp$ collisions} \label{pp-app} 
\setcounter{equation}{0}

\begin{figure}[htb]
\centerline{
  \scalebox{0.3}{\includegraphics{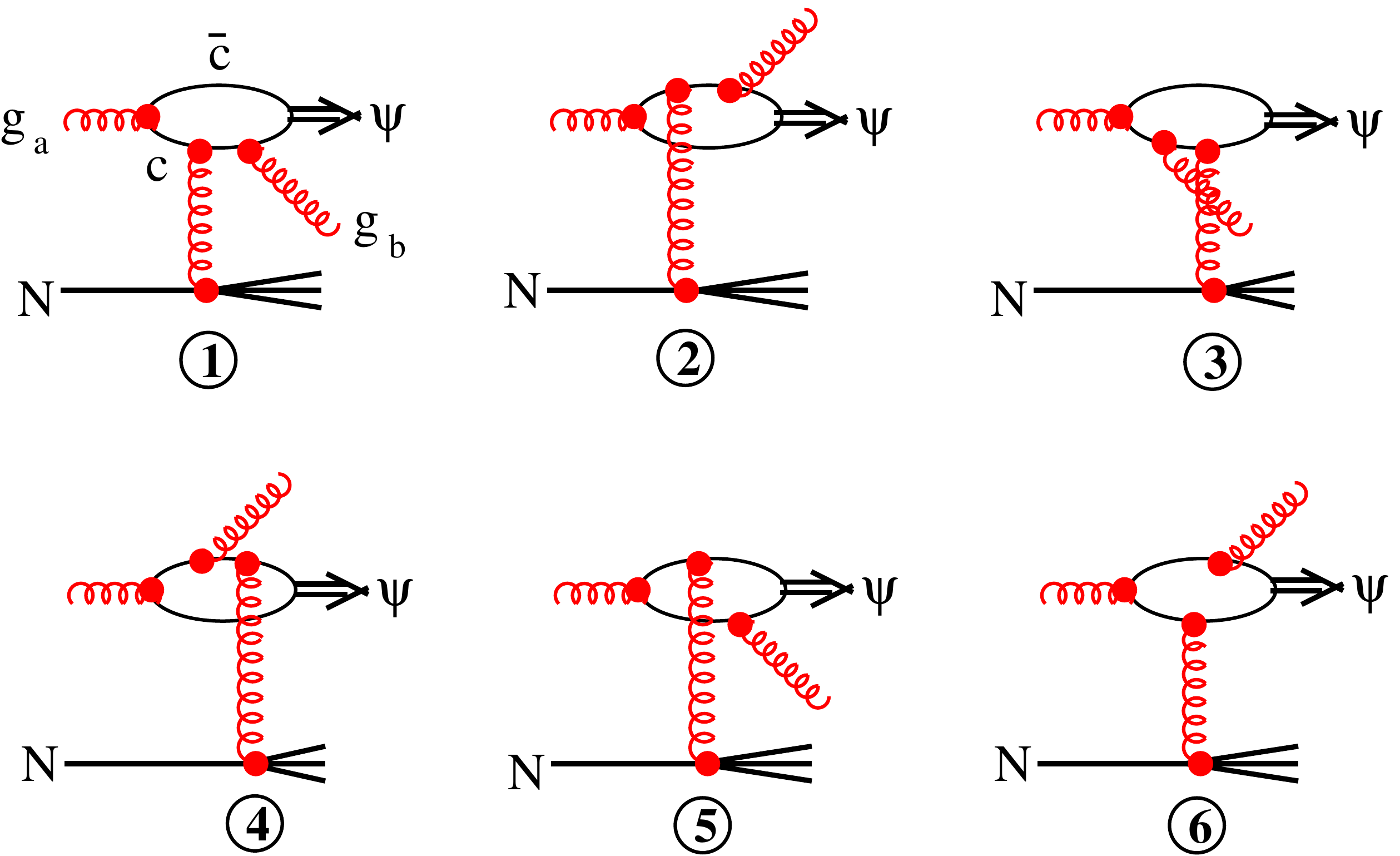}}}
\caption{\label{fig:6-graphs} (Color online) Feynman graphs for CSM of $J/\psi$ production.}
 \end{figure}

The production of heavy quarks was described within the dipole approach in \cite{kt-hf}. 
In the leading order of pQCD it is described by 15 Feynman graphs (Fig.~8 in \cite{kt-hf}).
Only six of them, presented here in Fig.~\ref{fig:6-graphs}, contribute to the production of 
$J/\psi$ and its excitations.

\subsection{Soft gluon approximation}

The amplitude, corresponding to these graphs was derived in \cite{kt-hf}, in the approximation of small fractional gluon LC momentum $\alpha_g\ll1$. 
\begin{widetext}
\beqn
&&\mathcal{A}_{abc}^{g_aN\to\psi g_bX}(\vec k_T,\vec k_g) = 
\frac{\sqrt{3}}{2}id_{abc}\int\limits_0^1 d\alpha \int d^2b\, d^2r\, d^2\rho\, \exp\left[i\vec k_g\cdot\vec \rho+i\vec k_T\cdot \vec b\right]
 \Psi_{\psi}\left(\alpha,\vec{r}\right)
 \nonumber \\ & \times & 
\left\{ 
 \Phi_{\bar{c}c}\left(\frac{\alpha}{1+\alpha_g},\left(1-\frac{\alpha_g}{\bar\alpha}\right)\vec{r}+\frac{\alpha_g}{\bar\alpha}\vec{\rho}\right)\Phi_{cg}\left(\frac{\vec{\rho}-\alpha\vec{r}}{\bar\alpha}\right)\gamma\left(\vec{b}+\frac{\left(\bar\alpha-\alpha_g\right)\alpha}{\bar\alpha}\,\vec{r}+\frac{\alpha_g}{\bar\alpha}\,\vec{\rho}\right)\right.\nonumber \\
 & - & \left.\Phi_{\bar{c}c}\left(\frac{\alpha+\alpha_g}{1+\alpha_g},\,\frac{\alpha\vec{r}-\alpha_g\vec{\rho}}{\alpha+\alpha_g}\right)\Phi_{cg}\left(\frac{\vec{\rho}+\left(\bar\alpha-\alpha_g\right)\vec{r}}{\alpha+\alpha_g}\right)\gamma\left(\vec{b}-\frac{\left(\bar\alpha-\alpha_g\right)\alpha}{\alpha+\alpha_g}\,\vec{r}+\frac{\alpha_g}{\alpha+\alpha_g}\,\vec{\rho}\right)\right\}.
 \label{b100}
\eeqn
\end{widetext}
Here $\alpha$ and $\bar\alpha=1-\alpha$ are the fractional light-cone momenta of the $\psi$, carried by the charm quark and antiquark, respectively. The relative transverse momentum and separation of $c$ and $\bar c$ are $\vec k$ and $\vec r$ respectively. 
We employ here the result of \cite{kt-hf} for the production of a colorless $\bar cc$ pair in $S$-wave, but projecting it to the charmonium light-cone wave function,
$\Psi_{\psi}\left(\alpha,\vec{r}\right)$, normalized as
\beq
\int\limits_0^1d\alpha\int d^2r\,
 \left|\Psi_{\psi}\left(\alpha,\vec{r}\right)\right|^2=1
\label{b120}
\eeq

The transverse momentum of $\psi$ as a whole, $p_{\psi}$, is related to the transverse momentum transfer to the target, $\vec k_T$, and the transverse momentum $\vec k_g$ of the radiated gluon as,
\beq
\vec k_T=\vec p_{\psi}+\vec k_g.
\label{b110}
\eeq
Further notations in (\ref{b100}) are the transverse distances $\vec b$ between the target and the center of gravity of $\psi$-$g$, and $\rho$ between the $\psi$ and radiated gluon. 

The light-cone distribution function for a quark, radiating a transversely polarized gluon with fractional momentum $\alpha_{cg}$, was derived in \cite{kst1},
\beq 
\Phi_{cg}(\tau,\rho_{cg})={1\over\pi}\,\sqrt{\frac{\alpha_{s}}{3}}\,
\xi_{\mu}^{\dagger}\,\hat{\mathcal{Q}}_{cg}\,\xi_{\bar{\mu}}\,
K_{0}\left(\tau\rho_{cg}\right),
\label{b300}
\eeq
where $\vec\rho_{cg}$ is the transverse separation between the final gluon and quark, 
and $\tau^2=(1-\alpha_{cg})m_g^2+\alpha_{cg}^2m_c^2$. Notice that the non-perturbative effects strongly affect this distribution function, leading to a significant reduction of the mean quark-gluon separation. The magnitude of this reduction is constrained by the observed suppression of diffractive gluon radiation \cite{kst2}, as well by many other processes \cite{spots}.
Here we rely on the perturbative form Eq.~(\ref{b300} of the distribution function, but introduce an effective gluon mass $m_g\approx 0.7\GeV$, which can be treated as a transverse mass of the gluon, which has a transverse motion enhanced by the non-perturbative effects.

The indices $\mu$ and $\bar{\mu}$
in~\ref{b300} are quark helicities before and after the gluon emission, and
$\xi_{\bar{\mu}}$ and $\xi_{\mu}^{\dagger}$ are
the spinors of the initial and final quarks respectively. The operator $\hat{\mathcal{Q}}_{cg}$ has the form \cite{kst1}
\beqn
\hat{\mathcal{Q}}_{cg}&=&im_c\alpha_{cg}^2\,
\vec e^{\,*}(\vec n\times\vec\sigma)+
\alpha_{cg}\, \vec e^{\,*}(\vec\sigma\times\vec\nabla)
\nonumber\\ &-&
i(2-\alpha_{cg})\,\vec e^{\,*}\vec\nabla.
\label{340}
\eeqn

The light-cone distribution function for the $g\to \bar cc$ transition is given by 
\beqn
&&\Phi_{\bar{c}c}(\epsilon,\vec R)=\frac{\sqrt{2\alpha_{s}}}{4\pi}\,
\xi_{\mu}^{\dagger}\,\hat{\mathcal{Q}}_{\bar cc}\,\xi_{\bar{\mu}}\,
K_{0}(\epsilon R),
\label{350}
\eeqn
where
\beq
\hat{\mathcal{Q}}_{\bar cc}=
m_{c}\vec{\sigma}\cdot\vec{e}_{i}+i\left(1-2\beta\right)\vec{\sigma}\cdot\vec{n}+\left(\vec{\sigma}\times\vec{e}_{i}\right)\cdot\vec\nabla_{R}.
\label{360}
\eeq
and
\beq
\epsilon^2=m_{c}^{2}-\beta(1-\beta)m_{g}^{2}
\label{370}
\eeq

The fractional momentum $\beta$ of the $c$ quark emerging from the incoming gluon (see Fig.~\ref{fig:6-graphs}) is different from that in the final state, due to gluon radiation by either $c$, or $\bar c$ quarks. Correspondingly, $\beta=\alpha/(1+\alpha_g)$, or $\beta=(\alpha+\alpha_g) /(1+\alpha_g)$, as one can see in (\ref{b100}). Gluon radiation also changes the $\bar cc$ separation $\vec R$, which is different from the final $\vec r$, as one can see in the argument of  $\Phi_{\bar{c}c}$ in Eq.~(\ref{b100}).
The $\bar cc$ distribution function contains proper convolution with
a Clebsch-Gordan coefficient $\left\langle 1M\left|\frac{1}{2}\bar{\mu}\frac{1}{2}\mu\right.\right\rangle$, where $M$ is the spin projection.

Following the definitions of \cite{kt-hf}, the function $\gamma(b)$ in (\ref{b100})
corresponds to the Fourier image of the dipole destruction amplitude,
which can also be treated as an “elastic” (color-exchange) gluon-nucleon scattering amplitude.
It is related to the dipole cross-section as
\beq
\sigma(r)=\int d^{2}b\,\left|\gamma\left(\vec{b}+\bar{\alpha}\vec{r}\right)-\gamma\left(\vec{b}-\alpha\vec{r}\right)\right|^{2}.
\label{b200}
\eeq

\subsection{The general case of arbitrary \boldmath$\alpha_g$}

A gluon, as a vector particle, is usually radiated at high energies with a small fractional momentum $\alpha_g\sim1/\ln(s)$. However, in the process under consideration, the transition of a $\bar cc$ pair from color-octet to singlet states, small $\alpha_g$ values are suppressed by color screening, and one should go beyond this approximation, Eq.~(\ref{b100}), and rely on the general form of the amplitude, 
where
\beqn&&\mathcal{A}_{abc}^{g_aN\to\psi g_bX}(\vec k_T,\vec k_g)
=\frac{\sqrt{3}}{2}id_{abc}\int d\alpha\, d^{2}b\, d^{2}r\, d^{2}\rho 
\nonumber\\ &\times&
\exp\left[i\vec k_{g}\cdot\vec \rho+i\vec k_{T}\cdot \vec b\right]\,\Psi_{\psi}\left(\alpha,\,\vec{r}\right)  
\label{b400} 
\\
 & \times & \sum_{n=1}^{6}\eta_{n}{\rm Tr}\left[\Lambda_{M}\,\Phi_{\bar{c}c}\left(\epsilon_{n},\,\vec{r}_{n}\right)\Phi_{cg}\left(\tau_{n},\,\vec{\rho}_{n}\right)\right]\gamma\left(\vec{b}_{n}\right).
\nonumber
\eeqn
The functions under the trace operation are here 2$\times$2
matrices in quark helicity space (helicity indices are dropped).
 The matrix $\Lambda_{M}$ contains the convolution of spinors with the 
Clebsch-Gordan coefficients from the wave function, 
\beqn
\Lambda_{M}^{\mu\bar{\mu}}&=&
\left\langle 1M\left|\frac{1}{2}\bar{\mu}\frac{1}{2}\mu\right.\right\rangle \xi_{\mu}\xi_{\bar{\mu}}^{\dagger}
\nonumber\\ &=&
\left(\frac{1+\sigma_{3}}{2},\,\frac{\sigma_{1}}{\sqrt{2}},\,\frac{1-\sigma_{3}}{2}\right)_{M=+1,0,-1}^{\mu\bar{\mu}},
\eeqn
where $\sigma_{i}$ are the Pauli matrices in helicity space.
 The multiplier
\begin{equation}
\eta_{l}=\{1,1,-1,-1,-\alpha_{G},-\alpha_{G}\}
\end{equation}
takes into account the ordering of $t_{a}$ matrices
and a numerical pre-factor. 

The functions $\Phi_{cg}(\tau_n,\rho_{n})$ and $\Phi_{\bar{c}c}(\epsilon_{n},\, r_{n})$ are defined in (\ref{b300}) and (\ref{350}) respectively. The contributions of different graphs 
depicted in Fig.~\ref{fig:6-graphs} to the amplitude are summed in Eq.~(\ref{b400}).
The fractional momenta $\alpha_n$ and $\beta_n$, as well as the transverse separations $\vec\rho_{n}$ and $\vec r_n$,
depend on the number of the corresponding graph in Fig.~\ref{fig:6-graphs}.

 It is assumed that at least one of the
quarks is on-shell. The parameters $\epsilon_{n},\,\tau_{n}$ as
well as arguments $r_{n},\, r_{G,n}$ for different diagrams (1-6)
in the Figure~\ref{fig:6-graphs} are given by 
\beqn
\epsilon_{1}^{2} & = & \epsilon_{3}^{2}=m_c^{2}-\left(\bar\alpha-\alpha_g\right)\left(\alpha+\alpha_g\right)m_g^{2}\nonumber\\
\tau_{1}^{2} & = & \tau_{5}^{2}=\left(\frac{\alpha_g}{\alpha+\alpha_g}\right)^{2}m_c^{2}+\frac{\alpha}{\alpha+\alpha_g}m_g^{2}\nonumber\\
\tau_{3}^{2} & = & \frac{\epsilon_{5}^{2}}{\left(\alpha+\alpha_g\right)^{2}}=\frac{\bar\alpha_g\left(\alpha_gm_c^{2}+\alpha\left(\bar\alpha-\alpha_g\right)m_g^{2}\right)}{\left(\bar\alpha-\alpha_g\right)\left(\alpha+\alpha_g\right)}\nonumber\\
\tau_{2}^{2} & = & \tau_{6}^{2}=\left(\frac{\alpha_g}{\bar\alpha}\right)^{2}m_c^{2}+\left(\frac{\bar\alpha-\alpha_g}{\bar\alpha}\right)m_g^{2}\nonumber\\
\tau_{4}^{2} & = & \frac{\epsilon_{6}^{2}}{\bar\alpha_g^{2}}=\frac{\bar\alpha\left(\alpha_gm_c^{2}+\alpha\left(\bar\alpha-\alpha_g\right)m_g^{2}\right)}{\alpha\,\bar\alpha_g}\nonumber\\
\epsilon_{2}^{2} & = & \epsilon_{4}^{2}=m_c^{2}-\bar\alpha\,\alpha \,m_g^{2}\nonumber\\
\vec{r}_{1} & = & \vec{r}_{3}=\vec{r}_{5}=\frac{\alpha\bar\alpha_g\,\vec{r}-\alpha_g\vec{\rho}}{\alpha\,\bar\alpha_g+\alpha_{g}},
\nonumber \\
\vec{r}_{2} & = & \vec{r}_{4}=\vec{r}_{6}=-\frac{\left(\bar\alpha-\alpha_{g}+\alpha\alpha_g\right)\vec{r}+\alpha_{g}\vec{\rho}}{\bar\alpha+\alpha\alpha_g},
\nonumber \\
\vec{\rho}_{1}& = &\vec{\rho}_{3}=\vec{\rho}_{5}=-\vec{\rho}-\left(\bar\alpha-\alpha_{g}+\alpha\alpha_g\right)\vec{r}\nonumber\\
\vec{\rho}_{2}& = &\vec{\rho}_{4}=\vec{\rho}_{6}=-\vec{\rho} + \alpha\,\bar\alpha_g\vec{r}\nonumber\\
\vec{b}_{1} & = & \vec{b}+\frac{\alpha_{g}\vec{\rho}-\alpha\,\bar\alpha_g(\bar\alpha-\alpha_{g}+\alpha\alpha_g)\vec{r}}{\alpha+\alpha_{g}-\alpha\alpha_g},
\nonumber \\
\vec{b}_{2}& = &\vec{b}+\frac{\alpha_{g}\vec{\rho}+\alpha\,\bar\alpha_g(\bar\alpha-\alpha_{g}+\alpha\alpha_g)\vec{r}}{\bar\alpha+\alpha\alpha_g},
\nonumber\\
\vec{b}_{3} & = & \vec{b}_{6}=\vec{b}-(\bar\alpha-\alpha_{g}+\alpha\alpha_g)\vec{r},
\nonumber \\
\vec{b}_{4}& = &\vec{b}_{5}=\vec{b}+\alpha\,\bar\alpha_g\vec{r},
\label{b450}
\eeqn
where $\bar\alpha_g=1-\alpha_g$.

The $p_{T}$-integrated differential cross-section of the inclusive charmonium production, which describes the distribution over $\rho$ and $r$, can  be expressed
in terms of the dipole cross-section,
\begin{widetext}
\begin{eqnarray}
&&\frac{d\sigma(pp\to \psi X)}{dy\,d^2\rho\,d^{2}r\,d^{2}r^\prime} = \frac{9}{8}\, g(x_{1})\int d\alpha_g\,
d\alpha\,d\alpha^\prime\,\Psi_{\psi}^{*}\left(\alpha,r\right)\Psi_{\psi}\left(\alpha^{\prime}, r^{\prime}\right)
\label{b500}\\
 & \times & \sum_{n,n'=1}^{6}\eta_{n}\eta_{n'}{\rm Tr}\left[\Lambda_{M}\Phi_{\bar{c}c}\left(\epsilon_{n},\vec{r}_{n}\right)\Phi_{cg}\left(\tau_{n},\vec{\rho}_{n}\right)\right]{\rm Tr}\left[\Lambda_{M}\Phi_{\bar{c}c}\left(\epsilon^\prime_{n'},\,\vec{r}_{n'}^{\,\prime}\right)\Phi_{cg}\left(\tau_{n'},\vec{\rho}_{n'}\right)\right]^{*}\sigma_{\bar qq}\left(\vec{b}_{n}-\vec{b}^{\,\prime}_{n'}\right),
 \nonumber 
\end{eqnarray}
\end{widetext}
where $y$ is the charmonium rapidity, and
\beq
x_{1,2}= \frac{\sqrt{M_{\psi}^2+p_T^2}}{\sqrt{s}}\,e^{\pm y}.
\label{b550}
\eeq
In the difference $\vec b_{n}-\vec b_{n'}^{\,\prime}$ in (\ref{b550}) the $b$ dependence cancels, so the dipole cross-section $\sigma_{\bar qq}$ in (\ref{b500}) is function of $\vec\rho$ and
$\vec r$. 

\begin{figure}[htb]
\centerline{
  \scalebox{0.35}{\includegraphics{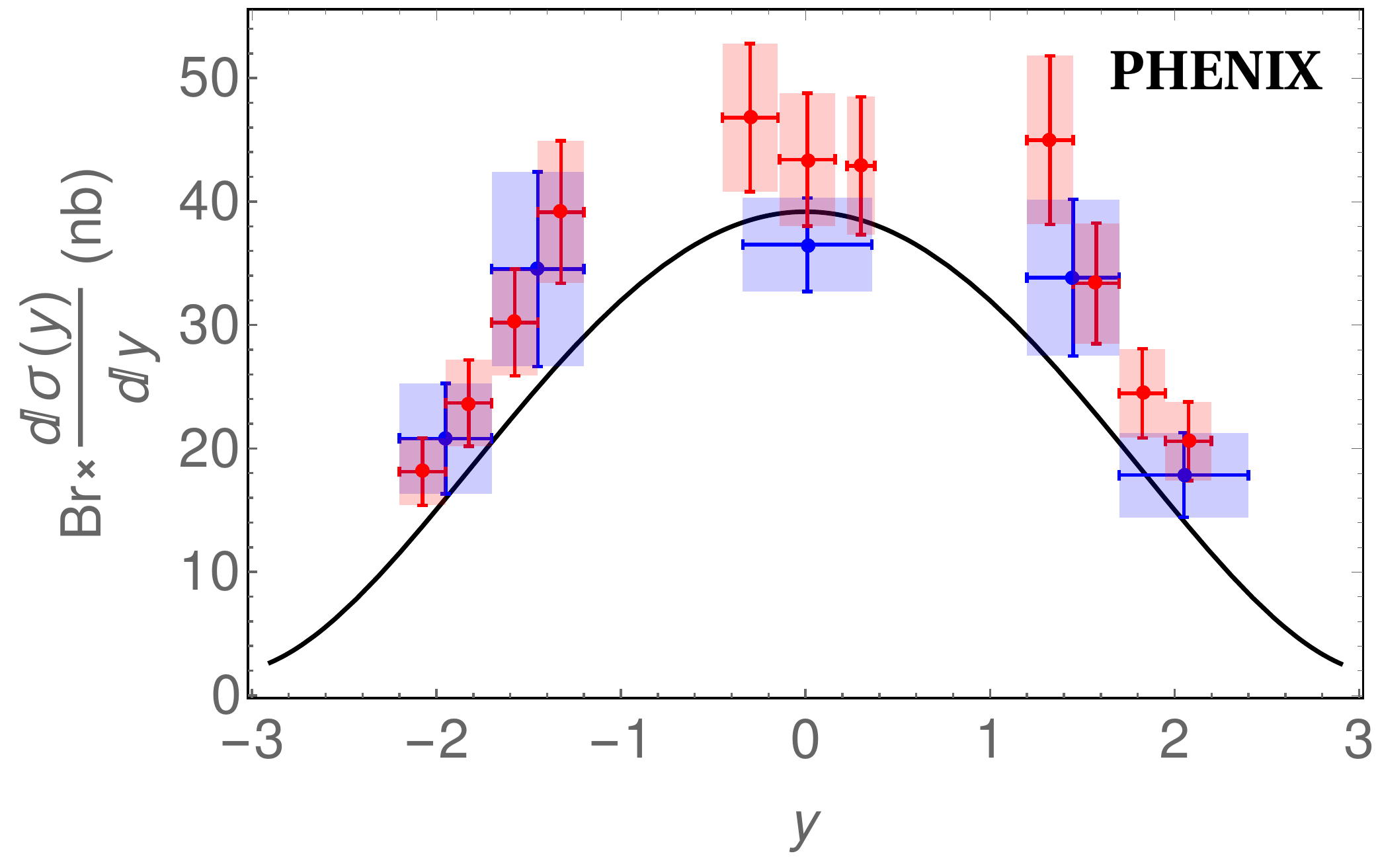}}}
\caption{\label{fig:pp-phenix} (Color online) The cross section of $pp\to J/\psi X$, calculated with (\ref{b500}), (\ref{b600}) in comparison with
data from \cite{phenix-pp,phenix-psi}  at $\sqrt{s}=200\GeV$.
}
 \end{figure}

\begin{figure}[htb]
\centerline{
  \scalebox{0.35}{\includegraphics{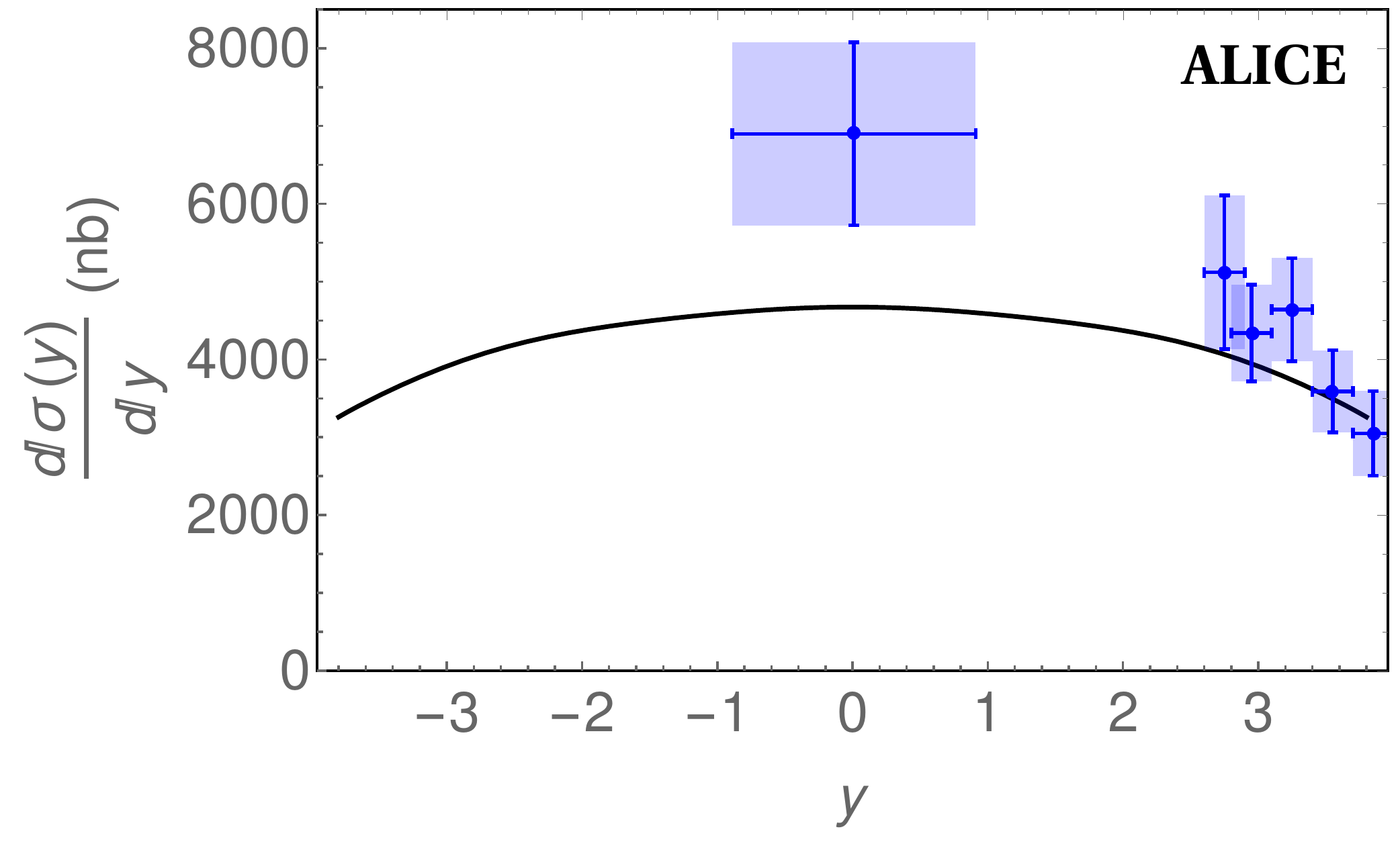}}}
\caption{\label{fig:pp-alice} (Color online) The same as in Fig.~\ref{fig:pp-phenix} at
$\sqrt{s}=5000\GeV$ in comparison with data \cite{alice-pp}.
}
 \end{figure}

The integrated cross section,
\beq
\frac{d\sigma(pp\to \psi X)}{dy} = 
\int d^2\rho\,d^{2}r\,d^{2}r^\prime\,
\frac{d\sigma(pp\to \psi X)}{dy\,d^2\rho\,d^{2}r\,d^{2}r^\prime},
\label{b600}
\eeq
can be compared directly with data. Comparison with available data from RHIC and LHC, is shown in Figs.~\ref{fig:pp-phenix} and \ref{fig:pp-alice}.

Although the calculations contain no free parameter adjusted to the data to be explained, there are theoretical uncertainties related to the different approximations that had been used. In particular, while the phenomenological dipole cross section takes into account the effects of gluon saturation, important at small $x_2$ in one of the protons, we rely on a single gluon approximation
in the projectile gluon distribution, which is justified only at large $x_1$.
Therefore the dipole description is "asymmetric",  it is reliable only at sufficiently small $x_2$,
but large $x_1$, and vice versa, i.e. at forward-backward rapidities, and in the central rapidity region is least reliable.

\end{document}